\documentclass[fleqn,usenatbib]{mnras}

\usepackage{newtxtext,newtxmath}

\usepackage[T1]{fontenc}

\DeclareRobustCommand{\VAN}[3]{#2}
\let\VANthebibliography\thebibliography
\def\thebibliography{\DeclareRobustCommand{\VAN}[3]{##3}\VANthebibliography}

\usepackage{graphicx}	
\usepackage{amsmath}	
\usepackage{lipsum}
\usepackage{ulem}
\usepackage{xcolor}
\definecolor{blazeorange}{rgb}{1.0, 0.4, 0.0}
\definecolor{seagreen}{rgb}{0.18, 0.55, 0.34}
\definecolor{rufous}{rgb}{0.66, 0.11, 0.03}
\definecolor{royalfuchsia}{rgb}{0.79, 0.17, 0.57}
\definecolor{scarlet}{rgb}{1.0, 0.13, 0.0}
\definecolor{royalpurple}{rgb}{0.47, 0.32, 0.66}
\definecolor{darkblue}{rgb}{0, 0, 0.66}

\newcommand\target{GRB\,220831A}
\newcommand\tninety{$T_{90}$}
\newcommand\epeak{$E_\mathrm{peak}$}
\newcommand\eiso{$E_\mathrm{iso}$}

\title[\target{}]{\target{}: a hostless, intermediate Gamma-ray burst with an unusual optical afterglow}

\author[Freeburn et al.]{
James Freeburn,$^{1,2}$\thanks{E-mail: jfreeburn@swin.edu.au}
Brendan O'Connor,$^3$\thanks{McWilliams Fellow}
Jeff Cooke,$^{1,2}$
Dougal Dobie,$^{2,4}$
Anais Möller,$^{1,2}$\newauthor
Nicolas Tejos,$^5$
Jielai Zhang,$^1$
Paz Beniamini,$^{6,7,8}$
Katie Auchettl,$^{2,9,10}$
James DeLaunay,$^{11}$\newauthor
Simone Dichiara,$^{11}$
Wen-fai Fong,$^{12}$
Simon Goode,$^{2,13}$
Alexa Gordon,$^{12}$
Charles D. Kilpatrick,$^{12}$\newauthor
Amy Lien,$^{14}$
Cassidy Mihalenko,$^{2,15}$
Geoffrey Ryan,$^{16}$
Karelle Siellez,$^{2,15}$
Mark Suhr,$^{1}$
Eleonora Troja$^{17}$,\newauthor
Natasha Van Bemmel,$^{1,2}$ and
Sara Webb$^{1,2}$
\\
$^1$Centre for Astrophysics and Supercomputing, Swinburne University of Technology, John St, Hawthorn, VIC 3122, Australia\\
$^2$ARC Centre of Excellence for Gravitational Wave Discovery (OzGrav), John St, Hawthorn, VIC 3122, Australia\\
$^3$McWilliams Center for Cosmology and Astrophysics, Department of Physics, Carnegie Mellon University, Pittsburgh, PA 15213, USA\\
$^4$Sydney Institute for Astronomy, School of Physics, University of Sydney, Sydney, NSW 2006, Australia\\
$^5$Instituto de F\'isica, Pontificia Universidad Cat\'olica de Valpara\'iso, Casilla 4059, Valpara\'iso, Chile\\
$^6$Department of Natural Sciences, The Open University of Israel, P.O Box 808, Ra’anana 43537, Israel\\
$^7$Astrophysics Research Center of the Open University (ARCO), The Open University of Israel, P.O Box 808, Ra’anana 43537, Israel\\
$^8$Department of Physics, The George Washington University, Washington, DC 20052, USA\\
$^9$School of Physics, The University of Melbourne, Parkville, VIC 3010, Australia \\ 
$^{10}$Department of Astronomy and Astrophysics, University of California, Santa Cruz, CA 95064, USA \\
$^{11}$Department of Astronomy and Astrophysics, The Pennsylvania State University, 525 Davey Lab, University Park, PA 16802, USA.\\
$^{12}$Center for Interdisciplinary Exploration and Research in Astrophysics (CIERA) and Department of Physics and Astronomy, Northwestern University,\\Evanston, IL 60208, USA\\
$^{13}$School of Physics and Astronomy, Monash University, VIC 3800, Australia\\
$^{14}$Department of Physics and Astronomy, University of Tampa, 401 W. Kennedy Blvd, Tampa, FL 33606, USA\\
$^{15}$School of Natural Sciences, University of Tasmania, Private Bag 37 Hobart, Tasmania, 7001, Australia\\
$^{16}$Perimeter Institute for Theoretical Physics, Waterloo, Ontario N2L 2Y5, Canada\\
$^{17}$University of Rome Tor Vergata, Department of Physics, via della Ricerca Scientifica 1, 00100, Rome, IT
}

\date{Accepted XXX. Received YYY; in original form ZZZ}

\pubyear{\the\year}

\begin{document}
\label{firstpage}
\pagerange{\pageref{firstpage}--\pageref{lastpage}}
\maketitle

\begin{abstract}

\target{} is a gamma-ray burst (GRB) with a duration and spectral peak energy that places it at the interface between the distribution of long-soft and short-hard GRBs. In this paper, we present the multi-wavelength follow-up campaign to \target{} and its optical, near-infrared, X-ray and radio counterparts.  Our deep optical and near-infrared observations do not reveal an underlying host galaxy, and establish that \target{} is observationally hostless to depth, $m_i\gtrsim26.6$ AB mag.  Based on the Amati relation and the non-detection of an accompanying supernova, we find that this GRB is most likely to have originated from a collapsar at $z>2$, but it could also possibly be a compact object merger at $z<0.4$ with a large separation distance from its host galaxy. Regardless of its origin, we show that its optical and near-infrared counterpart departs from the evolution expected from a forward shock dominated synchrotron afterglow, exhibiting a steep post-break temporal powerlaw index of $-3.83^{+0.62}_{-0.79}$, too steep to be the jet-break.  By analysing a range of models, we find that the observed steep departure from forward shock closure relations is likely due to an internal process producing either a flare or a plateau.

\end{abstract}




\section{Introduction}

Gamma-ray bursts (GRBs) have a bimodal distribution in their observed duration and spectral hardness \citep{two_classes}.  From this phenomenon, GRBs are traditionally divided into two classes.  Short-hard GRBs (SGRBs), with durations $\lesssim$2\,s, are associated with binary neutron star (BNS) mergers \citep{Eichler89,Gehrels05,at2017gfo5} while long-soft GRBs (LGRBs), with durations $\gtrsim$2\,s, are associated with collapsars \citep{Woosley1993,SN1998bw,Hjorth03}.  However, the duration and hardness of the prompt $\gamma$-ray emission do not necessarily determine a progenitor for a specific event  \citep{Zhang2009}.  There are two reasons for this:  Firstly, the distributions of SGRBs and LGRBs overlap significantly, preventing a secure classification of intermediate duration ($\sim$1-3 s) GRBs.  Secondly, there is an emerging population of GRBs, securely in the LGRB distribution that are associated with mergers \citep{Norris2006}.  

The discovery of a supernova (SN) following the $\sim1$\,s duration, GRB\,200826A \citep{200826A_ahumada,200826A_zhang,200826A_rossi,200826A_rhoades} exemplified the overlap between the two distributions \citep[e.g.,][]{Bromberg2012,Bromberg2013}. Kilonovae (KNe) are associated with the r-process nucleosynthesis that occurs following the merger of two neutron stars.  KNe were discovered associated with GRBs 211211A \citep{211211A_Troja,Rastinejad211211A,Yang211211A,Mei211211A,Gompertz211211A} and 230307A \citep{Levan230307A,Sun230307A,Yang230307A,Gillanders230307A,Dichiara230307A}, despite being firmly in the LGRB distribution with durations in excess of 50\,s \citep[though see][for alternative explanations]{Barnes23}.  These events cannot be explained merely in terms of scatter between the distributions of short and long duration GRBs, and conclusively demonstrate that the duration overlap between collapsars and mergers is larger than previously thought. The rest-frame \epeak{}--\eiso{} correlation, often called the Amati relation, along with other similar empirical correlations \citep[e.g.,][]{Guiriec_relation,Golenetskii_relation}, can be used to partially distinguish the two populations \citep{amati_relation1,amati_relation2}.  

Distinguishing between GRB progenitors can be aided by the identification of a host galaxy and a redshift determination. LGRBs trace star formation and often reside in star forming galaxies coincident with their host's stellar light distribution \citep{LGRB_SF_Bohdan,LGRB_SF_Bloom,Fruchter2006}, whereas short GRBs are found in a variety of different galaxies \citep{Leibler2010,natalkick,Fong2013,Nugent2022} and generally reside at larger distances from the brightest parts of their host \citep{natalkick,Fong2022,SGRB_hostgals}. Due to their differing redshift distributions and luminosities, arising from their distinct progenitor channels, the observed redshifts of these populations also differ significantly: LGRBs detected with \textit{Swift} have a mean redshift of 2.8 \citep{mean_z_GRB}, whereas SGRBs have a lower mean redshift of $\sim$0.6 \citep{SGRB_hostgals,Fong2022,Nugent2022}. 

GRB host galaxies can be difficult to identify and significant selection biases exist for both LGRBs \citep{LGRB_hosts} and SGRBs at $z$\,$>$\,$1$ \citep[e.g.,][]{SGRB_hostgals} due to the faintness of their host galaxies in this redshift range. Similarly, identifying some SGRB host galaxies can be complicated by a large angular offset, due potentially to natal kicks in the formation of BNS systems \citep{LGRB_SF_Bloom,hostlessBerger,natalkick,Behroozi2014,BP2024}.  It could also be due to long delay times between formation and merger, causing them to be associated with older and highly extended stellar populations in the host galaxy halos, that often remain unresolved \citep{PeretsBeniamini2021}. Indeed, a large fraction of the SGRB population ($\sim$30 percent) do not have a spatially coincident host galaxy despite deep imaging \citep[$\gtrsim$26 AB mag;][]{hostlessBerger,hostlessTunnicliffe,SGRB_hostgals}. For those SGRBs with identified hosts, their progenitors have projected distances of tens of kpc from their host \citep{hostlessBerger,natalkick,SGRB_hostgals,Fong2022}. 

\target{} was a GRB detected by the Fermi Gamma-ray Burst Monitor \citep[\textit{Fermi}/GBM;][]{fermi} and the \textit{Neil Gehrels Swift Observatory}'s Burst Alert Telescope \cite[\textit{Swift}/BAT;][]{swift} with a short duration ($\lesssim$2\,s) and a soft spectral profile \citep{WoodGCN,TohuvavohuGCN}.  This placed it in the intermediate region between SGRBs and LGRBs.  The afterglow to \target{} was detected at optical, near-infrared and radio wavelengths \citep{FreeeburnGCN1,DavanzoGCN,AndersonGCN} and upper limits were placed in X-rays \citep{DichiaraGCN}.  The optical and near-infrared afterglow was observed with the Dark Energy Camera (DECam), mounted on the Victor M. Blanco telescope, the \textit{Gemini} South telescope, and the Very Large Telescope (VLT), yielding an unusually steep post-break temporal decay, which deviates from forward shock closure relations.  \target{} is similar in its properties to GRB\,210704, which was an intermediate class burst, with a similarly difficult classification and deviations from a typical GRB afterglow \citep{Becerra23}.  Its peculiar properties highlight the importance of studying intermediate bursts like \target{}.

In this work, we investigate \target{}'s observational properties to shed light on its progenitor.  In Section \ref{sec:data}, we present observations of \target{}'s prompt emission and afterglow, spanning from $\gamma$-ray to radio wavelengths.  Without the identification of a host galaxy we evaluate the possible scenarios for \target{}'s host in Section \ref{sec:hostgalaxy}.  We fit both empirical and forward shock models to the multi-wavelength observations of \target{}.  We then assess the deviations from canonical models by fitting the optical and near-infrared (OIR) data with an additional emission component in the form of either a flare or an internal plateau.  This is presented in Section \ref{sec:multi_analysis}.  We then conclude that the likely progenitor of \target{} is a high-$z$ collapsar and interpret the afterglow's observed departures from closure relations in Section \ref{sec:discussion}.  Throughout this work we assume the cosmology reported in \citet{Planck18}, report all uncertainties at the 1$\sigma$ level and quote upper limits at the 5$\sigma$ level.

\section{Data}\label{sec:data}

\subsection{High energy facilities}

\begin{figure}
 \includegraphics[width=\columnwidth]{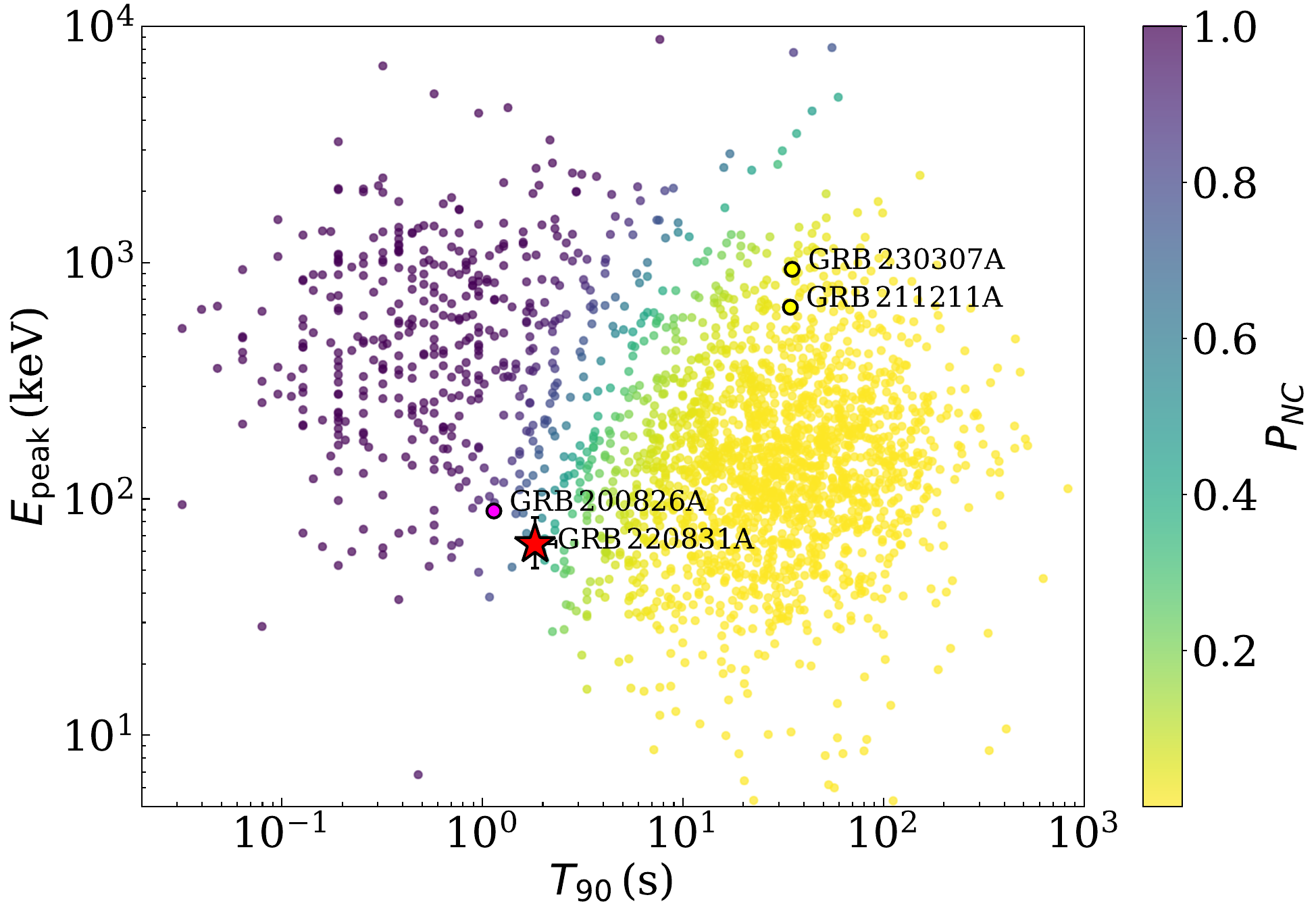}
 \caption{\tninety{} versus the observer frame spectral peak energy, \epeak{} from the \textit{Fermi}-GBM Gamma-Ray Burst Spectral Catalog \citep{GBMspectralcat} fit with the Band function \citep{Bandfunction}.  We fit the distribution with two Gaussian distributions in log-space. Based on this fit, each GRB is colour-coded with the probability of it being a non-collapsar, $P_{\mathrm{NC}}$ \citep[e.g.,][]{Bromberg2013}, where yellow and purple denote a 0 and 100 percent probability of being a non-collapsar respectively.}
 \label{fig:GRB_pop}
\end{figure}

\textit{Fermi}'s Gamma-ray Burst Monitor (GBM) detected \target{} at 13:56:32.93 UT on 31 Aug 2022 \citep{WoodGCN}, hereafter taken to be the start time, $T_0$, of the GRB.  We conducted analysis on the standard data products from the \textit{Fermi}/GBM burst catalog \citep{FermiBurstCat}.  \target{} was visible only in five of the twelve NaI detectors ($n_{\mathrm{a}}$, $n_{\mathrm{b}}$, $n_9$, $n_8$ and $n_7$) and neither of the BGO detectors that comprise \textit{Fermi}/GBM.  We calculate the $T_{90}$ for \target{}, defined by the difference in time between where 5 percent and 95 percent of \target{}'s fluence is emitted \citep{two_classes}. To do this, we utilise just the $n_{\mathrm{b}}$ detector, where the burst was brightest, and bin over 20--100\,keV.  We simulate a range of bursts based off the error associated with each time bin and generate a distribution of $T_{90}$ values to yield $T_{90}=1.83^{+0.50}_{-0.10}$\,s.  We performed a spectral analysis of the burst, from the five NaI detectors in which it is visible, using the \textit{Fermi} Gamma-ray Data Tools package \citep{GDT-Fermi} between 4--100\,keV and binning between $T_0-1.5$\,s and $T_0+1.5$\,s.  The burst is best-fit with a Band function, with the peak spectral energy, $E_{\mathrm{peak}}= 64^{+19}_{-14}$\,keV,  the low energy photon index, $\alpha_{\gamma} = -1.46^{+0.06}_{-0.07}$ and the high energy photon index, $\beta_{\gamma} = -3.8^{+6.2}_{-1.8}$.  We place these properties in the context of other \textit{Fermi}/GBM detected GRBs in the \textit{Fermi}/GBM Gamma-ray Burst Spectral Catalog \citep{GBMspectralcat} in Figure \ref{fig:GRB_pop} and analyse these results in Section \ref{sec:prompt_analysis}.

\textit{Swift}/BAT was not successfully triggered on-board by \target{}. No prompt alerts were provided by \textit{Fermi}/GBM due to communication issues. With no prompt alert, the Gamma-Ray Urgent Archiver for Novel Opportunities \citep[GUANO;][]{BATGUANO}, a pipeline used to archive \textit{Swift}/BAT event data around times of interest, was unable to be triggered.  Fortunately, there was a spurious transient $\sim$5\,s prior to the trigger time that tripped one of \textit{Swift}/BAT's many rate trigger algorithms. When no point source was found in the on-board image, this trigger was classified as a ``failed'' trigger, which \textit{Swift}/BAT archives 10\,s of event data. The 10 s cover 7.7\,s prior to the trigger time to 2.4\,s after trigger, including the full prompt emission period of the burst. The spurious transient is separated temporally from the burst emission and did not affect the analysis of the burst. Using these event data the \textit{Swift}/BAT-GUANO team was able to create a sky image and localize this burst with an uncertainty of 2\arcmin\, \citep{TohuvavohuGCN}. The initial analysis of the burst placed the measured duration of $T_{90}$ at $\sim$1\,s \citep{TohuvavohuGCN}, placing it in the SGRB regime \citep{two_classes}. Using the \textit{Swift}/BAT tools in the \textsc{HEASOFT} \citep{HEASOFT} software package, a spectral file was created from the event data over the full emission period of the burst (0.47\,s prior to trigger time to 1.13\,s after trigger time), as well as a detector response matrix. We fit the data with \textsc{XSPEC} \citep{xspec} and found $E_{\mathrm{peak}}=62\pm46$\,keV from a Band function fit. Using the \textsc{battblocks} tool we found a $T_{90}=1.3\pm0.3$\,s. 

\subsection{\textit{Swift} Follow-up Observations}

The X-ray Telescope (XRT) on-board \textit{Swift} observed \target{} starting on 1 November 2022 at 04:09:52 UT, corresponding to 0.6\,d after the initial \textit{Fermi} trigger. \citet{DichiaraGCN} report a low-significance detection at the location of the OIR afterglow (see Section \ref{sec:OIR_data}). To quantify the presence of X-ray emission, we reduce the two observations associated with this event (ObsID: 00021512001 and 00021512002). We reprocessed all observations from level one XRT data using \textsc{xrtpipeline} version 0.13.7. Here we use the most recent calibrations files and standard filter and screening criteria\footnote{\url{http://swift.gsfc.nasa.gov/analysis/xrt_swguide_v1_2.pdf}}. Using a source region with a radius of 49\arcsec\, centered on the position of \target{} and a source-free background region with a radius of 120\arcsec\, centered at $(\alpha,\delta)$=(01:37:11.2426, -41:31:30.093), we found no significant (>$3\sigma$) X-ray emission associated with individual observations that covered the GRB. To increase the signal to noise of our observations, we merged the two individual \textit{Swift} observations using \textsc{XSELECT} version 2.5b. Similarly, we detected no X-ray emission associated with the position of the \target{}, with a 0.3-10 keV 5$\sigma$ upper limit of 0.002\,counts\,s$^{-1}$. This suggests the possible detection by \citet{DichiaraGCN} was spurious. Assuming an absorbed powerlaw with a column density of $N_{H}=1.59\times10^{20}$\,cm$^{-2}$, a spectral index, $\beta$, of -0.69, measured from a fit to the OIR afterglow in Section \ref{sec:obs_prop} and a redshift of 2.4 (see Section \ref{sec:hostgalaxy}), we derive a 5$\sigma$ upper limit to the flux of $4.43\times10^{-13}$\,erg\,cm$^{-2}$\,s$^{-1}$.

The Ultra-violet and Optical Telescope (UVOT) on-board \textit{Swift} provided simultaneous observations to XRT.  \citet{KlinglerGCN} reported an upper limit of >$23.4$ AB mag with the UVOT \textit{white} filter at $\sim$18\,h post-burst.

\subsection{Optical and near-infrared observations}\label{sec:OIR_data}

Here we summarize the deep OIR observations of \target{} aimed at both its initial afterglow (Table \ref{tab:afterglowdata}) and late-time searches for its host galaxy (Table \ref{tab:hostgaluplims}). These include observations with VLT, DECam, and 11.1\,hr of Gemini-South observations across multiple programs.

\target{}'s OIR counterpart was discovered at $(\alpha,\delta)$=(01:37:00.95, -41:35:34.37), 19.4\,h post-burst with VLT/HAWK-I follow-up \citep[PI: Tanvir,][]{DavanzoGCN}. Subsequently, the optical counterpart was observed using DECam between 1 November 2022 and 4 November 2022, corresponding to 0.6\,d and 3.8\,d post-burst respectively \citep[PI: Cooke,][]{FreeeburnGCN1,FreeburnGCN2}.  An additional two epochs of VLT observations with FORS2 and HAWK-I were conducted at 1.7\,d and 2.8\,d post-burst.  Later, the optical counterpart was observed with \textit{Gemini}/GMOS at 4.6\,d post-burst \citep[PI: Gordon,][]{GordonGCN} in $r$ and $i$ filters, revealing a steep temporal break in the OIR light curve.

\textit{Gemini}/FLAMINGOS-2 near-infrared observations (PI: O'Connor) were carried out in the $J$ and $Ks$ filters at 9, 10 and 52\,d post-burst. These data  appeared to show a near-infrared source \citep{OconnorGCN}, spatially coincident with the OIR afterglow with a brightness of $\sim$24.5\,AB mag and 23.5\,AB mag, respectively. However, the counterpart, detected with \textit{Gemini}/FLAMINGOS-2, had a small angular offset from the previous OIR afterglow observations of $\sim$0.6\arcsec\, (see Figure \ref{fig:artefact} and Appendix \ref{sec:artefact}).  Moreover, VLT/HAWK-I observations at $\sim$9\,d post-burst, shortly after the initial \textit{Gemini} epoch, yielded an upper limit of $J\gtrsim24.9$ AB mag, which conflicts with the brightness of the multi-epoch \textit{Gemini} $J$-band detections. Based on a detailed re-analysis of all of the \textit{Gemini} OIR data, we conclude that the near-infrared detections were spurious and likely due to an unknown image artefact caused by residuals from the sky subtraction combined with the dither pattern (Kathleen Labrie, private communication). The artefact was not present in the late-time (July 2023) $J$-band Gemini observations discussed below. A more detailed analysis of these observations is included in Appendix \ref{sec:artefact}.

After the OIR counterpart to \target{} had faded, the field was subsequently observed using VLT HAWK-I and FORS2 instruments and provided upper limits at 5.7, 8.7, 22.6 and 26.8 days post-burst in the $R$, $J$, $R$, and $J$ filters, respectively \citep[PI: Tanvir,][]{DavanzoGCN}.  It was also observed by \textit{Gemini}/GMOS at 24.5\,d post-burst in $Z$-band (PI: O'Connor).  Further late-time, deep observations were conducted with \textit{Gemini} in $J$-band in July 2023 (PI: O'Connor) and in the $g$, $i$, $J$ and $K_s$ filters between December 2023 and March 2024 (PI: Tejos).  These observations did not reveal any likely host galaxy coincident with the subarcsecond OIR afterglow localization. The upper limits on an angularly coincident host galaxy are provided in Table \ref{tab:hostgaluplims}.

The photometric measurements presented in Tables \ref{tab:afterglowdata} and \ref{tab:hostgaluplims} were conducted with aperture photometry using \textsc{Source Extractor} \citep{sextractor}.  For filters $g$, $R$, $r$, $i$, $I$ and $Z$, the photometric catalogue from the Dark Energy Survey's second data release \citep{DES_DR2} was used for calibration with Lupton transformations\footnote{\url{http://www.sdss3.org/dr8/algorithms/sdssUBVRITransform.php}} for $R$ and $I$.

\subsection{Radio observations}

\target{} was observed with the Australia Telescope Compact Array (ATCA) at 4 and 11\,d post-burst by an independent group \citep[PI: Anderson,][]{AndersonGCN}. Both observations were carried out using the 4\,cm receiver, with two 2048\,MHz continuum bands centered on 5.5 and 9\,GHz. We have calibrated and imaged the publicly available data from both observations using the standard {\sc miriad} techniques. We report a marginal detection of $55 \pm 15\,\mu$Jy at 9\,GHz in the second observation, but no detections otherwise (see Table \ref{tab:afterglowdata} for details).

\begin{table}
    \begin{center}        
    \caption{Observations of \target{}'s afterglow.}
    \label{tab:afterglowdata}
    \begin{tabular}{ccccc}
        \hline\hline
        Instrument & Band & Time & Flux Density & Ref.\\
                   &      &  Days after GRB & $\mu$Jy &\\
        \hline
        \textit{Swift}/ & 1 keV & 0.75 & $<0.096$ & 1\\
        XRT            &       & &        & \\
        \hline
        \textit{Swift}/ & \textit{white} & 0.75 & $<1.6$ & 2\\
        UVOT            &       & &        & \\
        \hline
        \textit{Blanco}/ & $g$ & 0.8 & $1.39\pm0.19$ & 3,4\\
        DECam & $r$ & 0.8 & $1.94\pm0.14$ & \\
        & $i$ & 0.8 & $2.17\pm0.18$ & \\
        & $g$ & 1.6 & $0.82\pm0.18$ & \\
        & $r$ & 1.6 & $0.90\pm0.13$ & \\
        & $i$ & 1.6 & $1.22\pm0.11$ & \\
        & $g$ & 1.8 & $<0.96$ & \\
        & $r$ & 1.8 & $0.93\pm0.14$ & \\
        & $i$ & 1.8 & $1.09\pm0.12$ & \\
        & $i$ & 3.8 & $<0.62$ & \\
        \hline
        VLT/ & $J$ & 0.8 & $2.75\pm0.23$ & 5\\
        HAWK-I & $J$ & 1.7 & $1.500\pm0.097$ & \\
         & $J$ & 2.7 & $0.85\pm0.11$ & \\
         & $K_s$ & 2.8 & $1.23\pm0.16$ & \\
         & $J$ & 26.8 & $<0.74$ & \\
        \hline
        VLT/ & $R$ & 2.7 & $0.331\pm0.022$ & \\
        FORS2 & $I$ & 2.7 & $0.501\pm0.044$ &\\
         & $R$ & 22.6 & $<0.21$ & \\
        \hline
        ATCA & 9\,GHz & 4.1 & $<70$ & 6\\
        & 5.5\,GHz & 4.1 & $<70$ & \\
        & 9\,GHz & 11.2 & $55\pm15$ & \\
        & 5.5\,GHz & 11.2 & $<45$ & \\
        \hline
        \textit{Gemini}/ & $r$ & 4.6 & $<0.12$ & 7\\
        GMOS & $i$ & 4.6 & $0.109\pm0.031$ & \\ 
         & $Z$ & 57.4 & $<0.58$ & \\
        \hline\hline
    \end{tabular}
    \end{center}
    \footnotesize{\textbf{References.} (1) \cite{DichiaraGCN}; (2) \cite{KlinglerGCN}; (3) \cite{FreeeburnGCN1}; (4) \cite{FreeburnGCN2}; (5) \cite{DavanzoGCN}; (6) \cite{AndersonGCN}; (7) \cite{GordonGCN}.}
\end{table}

\begin{table}
    \caption{Brightness constraints on a host galaxy spatially coincident with \target{}.}
    \label{tab:hostgaluplims}
    \centering
    \begin{tabular}{cccc}
        \hline\hline
        Instrument & Band & $m$ & $M$ at $z=2.4$ \\
        \hline
        \textit{Gemini}/GMOS        & $g$ & $>26.2$ & $> -19.0$\\
                                    & $r$ & $>26.2$ & $> -19.0$\\
                                    & $i$ & $>26.5$ & $> -18.7$\\
                                    & $Z$ & $>24.5$ & $> -20.7$\\
        \hline
        VLT/FORS2                   & $R$ & $>25.6$ & $> -19.6$\\
        \hline
        VLT/HAWK-I                  & $J$ & $>24.9$ & $> -20.3$\\
        \hline
        \textit{Gemini}/FLAMINGOS-2 & $J$ & $>24.6$ & $> -20.6$\\
                                    & $K_s$ & $>24.0$ & $> -21.2$\\
        \hline\hline
    \end{tabular}
\end{table}

\section{Multi-wavelength Analysis}\label{sec:multi_analysis}

\subsection{A deep search for the host galaxy}\label{sec:hostgalaxy}

Despite deep follow-up observations with VLT and \textit{Gemini}, we do not detect a host galaxy within 2\arcsec\, of the subarcsecond localization of \target{} to $\gtrsim$ 26 AB mag (Table \ref{tab:hostgaluplims}). Therefore, we expand our search for a host galaxy to larger angular distances.

\subsubsection{NGC 625}

NGC 625 has an angular separation of 23.7\arcmin\, from \target{} and has a measured distance of $3.89\pm0.39$\,Mpc from Earth \citep{ngc625}.  \target{} would therefore lie $26.8\pm2.7$\,kpc from NGC 625 in projection, well within observed projected separations for SGRBs \citep{SGRB_hostgals,Fong2022}.  However, this would make \target{} the closest GRB ever discovered, whereas the brightness of its prompt emission would imply $E_{\gamma,\mathrm{iso}}=1.23^{+0.32}_{-0.20}\times10^{45}$\,erg, which is significantly less energetic than the observed, on-axis GRB population. At such a nearby distance, the continued deep monitoring by Gemini and VLT would have uncovered either a clear KN, assuming a brightness similar to AT2017gfo \citep{at2017gfo1,at2017gfo2,at2017gfo3,at2017gfo4,at2017gfo5,at2017gfo6,at2017gfo7}, or SN in excess of the afterglow (see Section \ref{sec:sec_emission}). The lack of these features implies a much larger distance than 4 Mpc. 

A magnetar giant flare (MGF) would be consistent with the isotropic-equivalent gamma-ray energy release at this distance and the lack of KN and SN detection.  Observations of Galactic magnetars show they are predominantly confined to the thin disk \citep{magnetars}.  \target{}'s location $26.8\pm2.7$\,kpc from NGC 625 would be strange for a magnetar and inconsistent with their magnetic field decay times of $10^3-10^4$\,yr \citep{Colpi00,Beniamini19}.  We therefore conclude that NGC 625 is an unlikely host galaxy for \target{}.

\subsubsection{Other field galaxies}

\citet{LGRB_SF_Bloom} calculate the probability of a transient being coincident with a given galaxy by chance, $P_{\mathrm{cc,gal}}$, using,
\begin{equation}
    P_{\mathrm{cc,gal}} = 1 - \exp[-\pi r_{\mathrm{gal}}^{2} \sigma(\leq m_{R,\mathrm{gal}})]
\end{equation}
where $r_{\mathrm{gal}}$ is the angular separation between the event and a given galaxy, $\sigma$ is the number density of galaxies at or brighter than the apparent magnitude, $m_{R,\mathrm{gal}}$, of the candidate host galaxy in $R$-band. In what follows, we utilize this probability of chance coincidence to test the likelihood of association for the surrounding galaxies in our deep imaging.

\begin{figure}
 \includegraphics[width=\columnwidth]{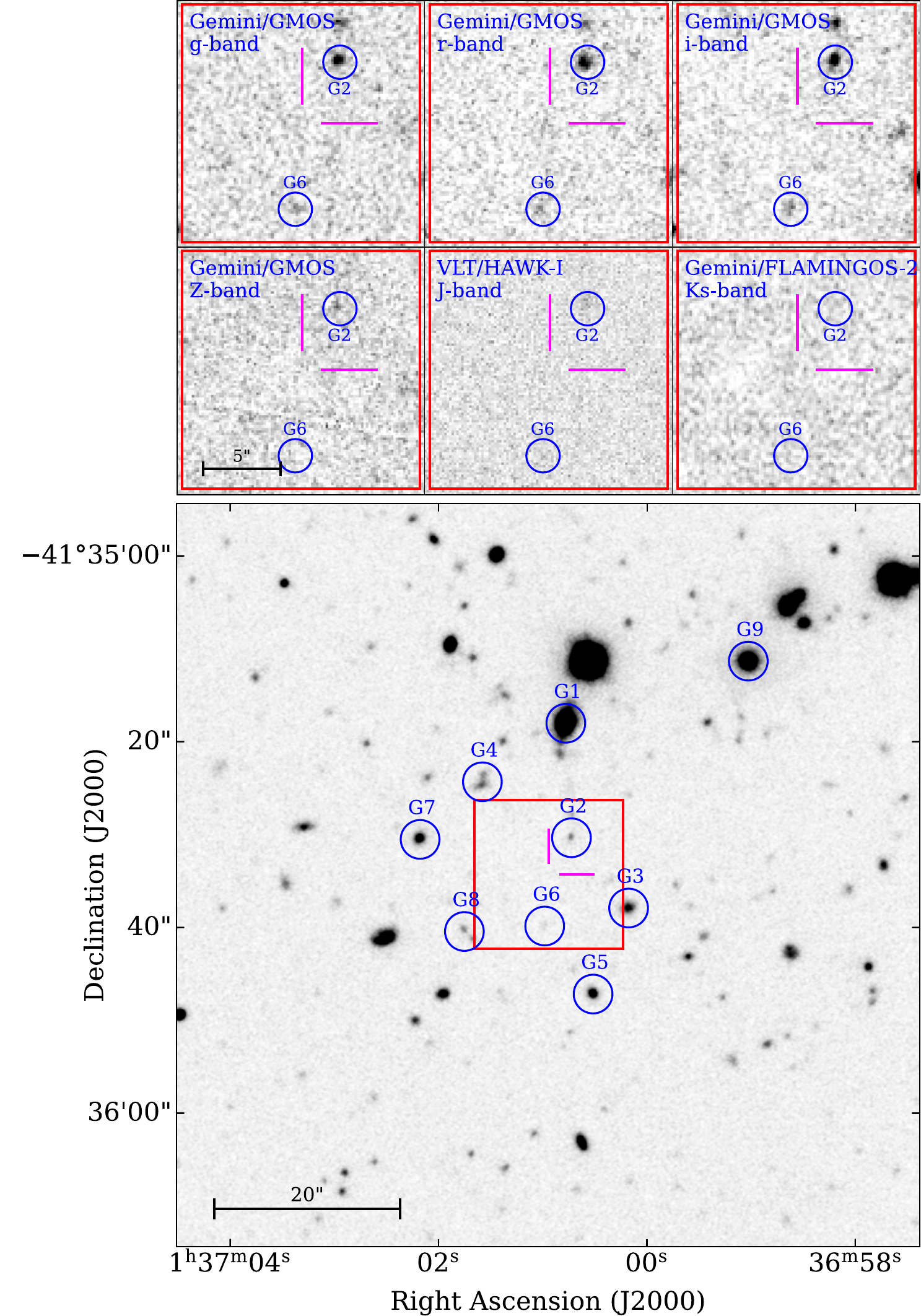}
 \caption{The bottom panel shows a \textit{Gemini}/GMOS $i$-band image of the field surrounding \target{}'s OIR afterglow with 5$\sigma$ depth of 26.5 AB mag.  The location of \target{}'s OIR afterglow is marked with the magenta cross hairs and galaxies within 1 arcminute with $P_{\mathrm{cc,gal}}<0.99$ are marked with blue circles and labelled G1-8.  Their properties are listed in Table \ref{tab:hostgalcands}.  There was no photometric redshift available for G2, G6 and G8. The top panel shows the 16\arcsec\, red box in the top plot, zoomed in and shown in different filters with which deep imaging was conducted.  The limits of these images are shown in Table \ref{tab:hostgaluplims}.}
 \label{fig:hostgal_field}
\end{figure}

Following \citet{SGRB_hostgals}, we consider a one arcminute radius from \target{}'s OIR sub-arcsecond afterglow localization in \textit{Gemini}/GMOS $i$-band imaging. This is because, for the vast majority of SGRBs, a confident host galaxy association has not been possible beyond a separation of 1\arcmin. We identify potential host galaxies by using our deep $r$ and $i$-band \textit{Gemini}/GMOS imaging and calculating an $R$-band magnitude with a Lupton transformation.  We then select those that have a 5$\sigma$ detection, satisfy $P_{\mathrm{cc,gal}}<0.99$ and a cut using \textsc{Source Extractor}'s star-galaxy classifier, \textsc{CLASS\_STAR} < 0.5.  The resultant selection of galaxies are shown in Figure \ref{fig:hostgal_field}. The position of these galaxies, in addition to their $R$-band AB magnitudes, photometric redshifts and projected physical offsets, are also listed in in Table \ref{tab:hostgalcands}. We find that G1, at an angular offset of 16.5\arcsec, has the lowest probability of chance coincidence at $P_{\mathrm{cc,gal}}=0.42$, which is a very unlikely host for \target{}. However, of the galaxies with measured photometric redshifts listed in Table \ref{tab:hostgalcands}, G1, G3 and G4 have separations that are consistent with the observed separation distances of SGRB host galaxies, which range up to $\sim$75\,kpc \citep{SGRB_hostgals,Fong2022}. While we cannot rule out the low-$z$ SGRB scenario, we disfavour it based on the lack of significant association with any of the galaxies shown in Table \ref{tab:hostgalcands}. In addition, because we find no coincident host galaxy down to the depths listed in Table \ref{tab:hostgaluplims}, we therefore conclude that \target{}'s host galaxy is not present in our observations. 

\subsubsection{An unseen host galaxy}

LGRBs are usually confined to star formation regions within their host galaxies \citep{LGRB_SF_Bloom,LGRB_SF_Bohdan}. A high-$z$ host not detected to the absolute magnitude limits shown in Table \ref{tab:hostgaluplims} lies in the $\sim80^{\mathrm{th}}$ percentile of the brightness of observed LGRB host galaxies for $1.9 < z < 2.7$ \citep{SchulzeGRBhosts}.  For \target{}, this would imply a range of $2.11 \times 10^{51}\mbox{\,erg} < E_{\gamma,\mathrm{iso}} < 5.49 \times 10^{51}\mbox{\,erg}$. This, along with its location on the Amati relation in Figure \ref{fig:amati_relation}, favours a faint, high-$z$ host underlying the position of \target{} that is not detected in our deep ground-based imaging.

\begin{table*}
    \caption{Host galaxy candidates for \target{}.}
    \label{tab:hostgalcands}
    \centering
    \begin{tabular}{ccccccc}
        \hline\hline
        Label & Coordinates & $m_{R,\mathrm{gal}}$ & Angular Separation & $P_{\mathrm{cc,gal}}$ & $z_{\mathrm{phot}}$ & Projected Distance \\
              &             & AB mag & \arcsec         &                   &                     & kpc \\
        \hline
        \vspace{0.02\columnwidth}
        G1 & 01:37:00.8 -41:35:18.0 & $21.452 \pm 0.011$ & 16.5 & 0.42 & $0.23\pm0.13$ & $61^{+22}_{-22}$ \\
        \vspace{0.02\columnwidth}
        G2 & 01:37:00.7 -41:35:30.3 & $24.585 \pm 0.064$ & 4.8 & 0.43 & -- & -- \\
        \vspace{0.02\columnwidth}
        G3 & 01:37:00.2 -41:35:37.9 & $23.660 \pm 0.036$ & 9.3 & 0.69 & $0.89\pm0.23$ & $72^{+4}_{-8}$ \\
        \vspace{0.02\columnwidth}
        G4 & 01:37:01.6 -41:35:24.3 & $23.870 \pm 0.039$ & 12.3 & 0.83 & $0.77\pm0.49$ & $93^{+12}_{-40}$ \\
        \vspace{0.02\columnwidth}
        G5 & 01:37:00.5 -41:35:47.1 & $23.375 \pm 0.029$ & 13.6 & 0.84 & $0.86\pm0.30$ & $107^{+8}_{-17}$ \\
        \vspace{0.02\columnwidth}
        G6 & 01:37:01.0 -41:35:39.9 & $25.91 \pm 0.19$ & 5.9 & 0.89 & -- & -- \\
        \vspace{0.02\columnwidth}
        G7 & 01:37:02.2 -41:35:30.3 & $23.792 \pm 0.044$ & 14.4 & 0.89 & $1.07\pm0.16$ & $120^{+3}_{-5}$ \\
        \vspace{0.02\columnwidth}
        G8 & 01:37:01.8 -41:35:40.4 & $24.870 \pm 0.083$ & 10.4 & 0.95 & -- & -- \\
        G9 & 01:36:59.0 -41:35:11.3 & $22.199 \pm 0.017$ & 31.5 & 0.97 & $0.85\pm0.06$ & $245^{+5}_{-6}$ \\
        \hline\hline
    \end{tabular}
\end{table*}

\subsection{The prompt emission}\label{sec:prompt_analysis}

Here we consider the prompt emission properties of \target{} in order to aid in determining its nature as either an SGRB or LGRB. To start, we fit the distribution of bursts from the \textit{Fermi}-GBM Gamma-Ray Burst Spectral Catalog \citep{GBMspectralcat} in \tninety{}--\epeak{} space with a two-dimensional bimodal distribution.  Figure \ref{fig:GRB_pop} shows that \target{} occupies an intermediate region in \tninety{}--\epeak{} space between SGRBs and LGRBs. Based on our fitted distributions and the \textit{Fermi}/GBM of \target{}, we obtain $P_{\mathrm{NC}}=0.54^{+0.10}_{-0.23}$.  Moreover, from the \textit{Fermi}/GBM data, with the procedure described in \citet{Bromberg2013}, we calculate the probability of \target{} being a non-collapsar as  $P_{\mathrm{NC}}=0.48^{+0.12}_{-0.34}$. These two methods are in agreement. Therefore, using the observer frame prompt emission properties alone, we cannot discern its nature due to its intermediate position in Figure \ref{fig:GRB_pop}. 

With the rest-frame spectral peak energy and the $\gamma$-ray, isotropic-equivalent energy release, $E_{\mathrm{rest,peak}}$--$E_{\gamma,\mathrm{iso}}$, the Amati relation can be used to infer the likely progenitor of a given GRB \citep{amati_relation1,amati_relation2}.  However, without a redshift, we cannot calculate the rest frame properties of \target{}.

\begin{figure}
 \includegraphics[width=\columnwidth]{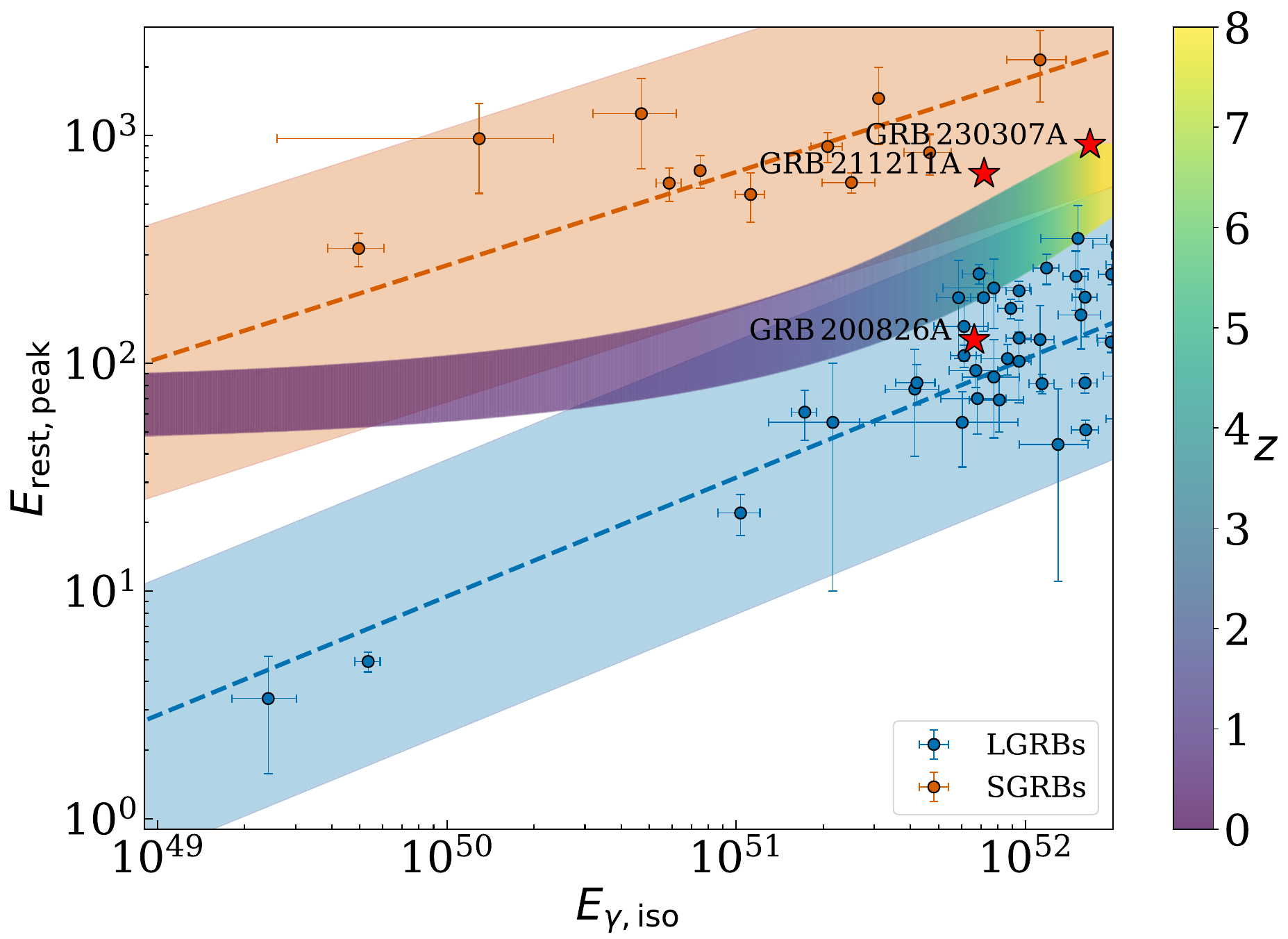}
 \caption{\target{} and a selection of SGRBs and LGRBs plotted in rest-frame $E_{\mathrm{rest,peak}}$--$E_{\gamma,\mathrm{iso}}$ space compared to the Amati relation \citep[adapted from][]{Dichiara_amati_plot}.  The Amati relation for SGRBs (orange) and LGRBs (blue) is plotted with a dotted line with a filled, shaded region denoting the 3$\sigma$ scatter in the correlation.  We plot \target{}'s location in $E_{\mathrm{rest,peak}}$--$E_{\gamma,\mathrm{iso}}$ space for $0<z<8$, where the colour bar denotes $z$.}
 \label{fig:amati_relation}
\end{figure}

Assuming a collapsar scenario for \target{}, we can use the $E_{\mathrm{rest,peak}}$--$E_{\gamma,\mathrm{iso}}$ correlation to calculate the redshift at which the emission properties of \target{} pass closest to the Amati relation (see Figure \ref{fig:amati_relation}).  This is called the `pseudo-redshift' \citep{amati_relation2} and we calculate a value of of $z=2.4$.  Similarly, for $z<0.4$, \target{} is consistent with the SGRB Amati relation to within 3$\sigma$ in $E_{\mathrm{rest,peak}}$--$E_{\gamma,\mathrm{iso}}$ space.  G1 and G4 in Table \ref{tab:hostgalcands} are consistent with this redshift range.

\subsection{The afterglow}\label{sec:multi_afterglow}

\subsubsection{Observed properties}\label{sec:obs_prop}

\begin{figure*}
    \includegraphics[width=0.766\textwidth]{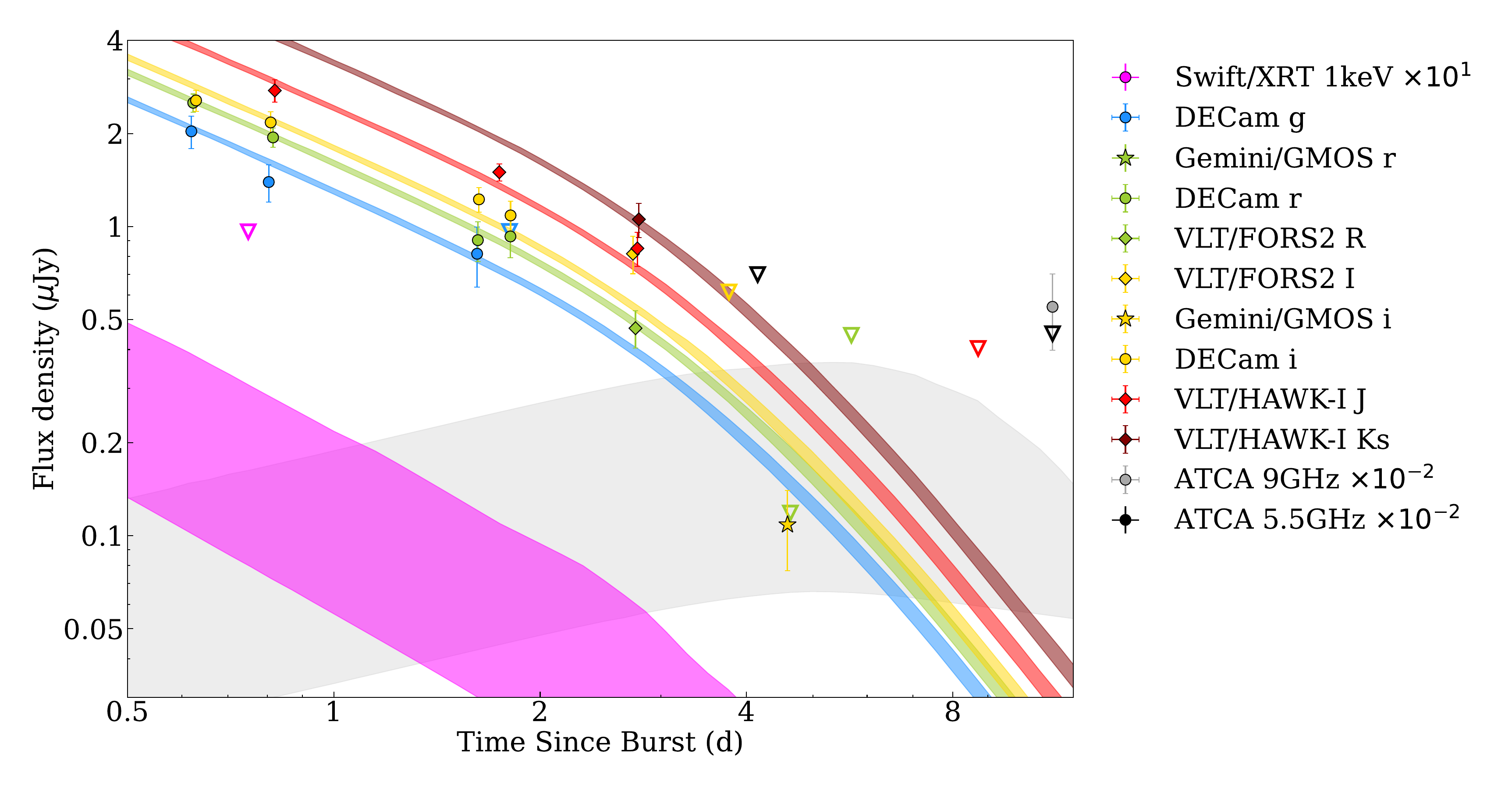}
    \caption{The light curve of \target{}'s afterglow. The shaded regions show the 1$\sigma$ intervals for the multi-wavelength, forward shock fit \textsc{afterglowpy} \citep{afterglowpy} model, assuming $z=2.4$ (the first row in Table \ref{tab:forwardshock_params}).  The X-ray and radio data and models have been rescaled for clarity.}
    \label{fig:lightcurve}
\end{figure*}

The OIR data, shown in Figure \ref{fig:lightcurve}, shows a clear temporal break before the last epoch of \textit{Gemini}/GMOS observations.  We fit the OIR and \textit{Swift}/XRT data with a broken powerlaw (BPL) of the form,
\begin{equation}
F(\nu,t) = A\nu^{\beta}
    \begin{cases}
        (t/t_b)^{\alpha_1}, & t<t_b \\
        (t/t_b)^{\alpha_2}, & t>t_b
    \end{cases}
\end{equation}
where A is the flux density scaling factor, $t$ is the time post-burst in days, $t_b$ is the time post-burst of a break in the temporal powerlaw index, $\nu$ is frequency, $\beta$ is the spectral index of the afterglow emission and $\alpha_1$ and $\alpha_2$ are the pre and post-break temporal powerlaw indices.

We explore the parameter space using \textsc{emcee} \citep{emcee}.  This yielded the following results; $t_b=2.60^{+0.10}_{-0.11}$\,d, $\alpha_1=-0.89^{+0.07}_{-0.07}$, $\alpha_2=-3.83^{+0.62}_{-0.79}$ and $\beta=-0.64^{+0.07}_{-0.07}$ with a reduced $\chi^2$ value of 0.75.  We therefore conclude that the afterglow data is consistent with a constant spectral index and a dramatic steepening of the fade rate between approximately 2 and 4 days post-burst.  

To ensure that this steepening is not sensitive to the sharpness of the break, we also fit a smoothly broken powerlaw of the form,
\begin{equation}
    \label{eq:sbpl}
    F(\nu,t) = A\nu^{\beta}\tau^{-\alpha_1} \left\{ \frac{1}{2} \left( 1 + \tau^{1/\Delta} \right) \right\}^{(\alpha_1 - \alpha_2)\Delta}
\end{equation}
where $\tau = (t-t_0)/(t_b-t_0)$, $t_0$ is the start time of the flare and $\Delta$ quantifies the smoothness of the break, which was left as a free parameter.  For the post-break power-law index, we obtained $\alpha_2=4.50^{+1.15}_{0.99}$. We therefore conservatively adopt the results of the sharp BPL fit to analyse the afterglow's agreement with closure relations.

Under the standard afterglow model wherein synchrotron emission results from a relativistic shock interacting with a uniform interstellar medium and observing the afterglow between the typical frequency and the cooling frequency, $\nu_m < \nu < \nu_c$, the electron energy distribution powerlaw index, $p$, predicts values of $\alpha = 3(1-p)/4$, $\beta = (1-p)/2$ for the pre-break light curve \citep{afterglow_spectra}.  Due to the lateral spreading of the jet, there is expected to be a steepening in the fade rate of the afterglow emission.  This is called the `jet-break' and is characterised by the temporal powerlaw index, $\alpha$, asymptoting to $-p$ \citep{Rhoads97,jetbreak}.  

\begin{figure}
 \includegraphics[width=\columnwidth]{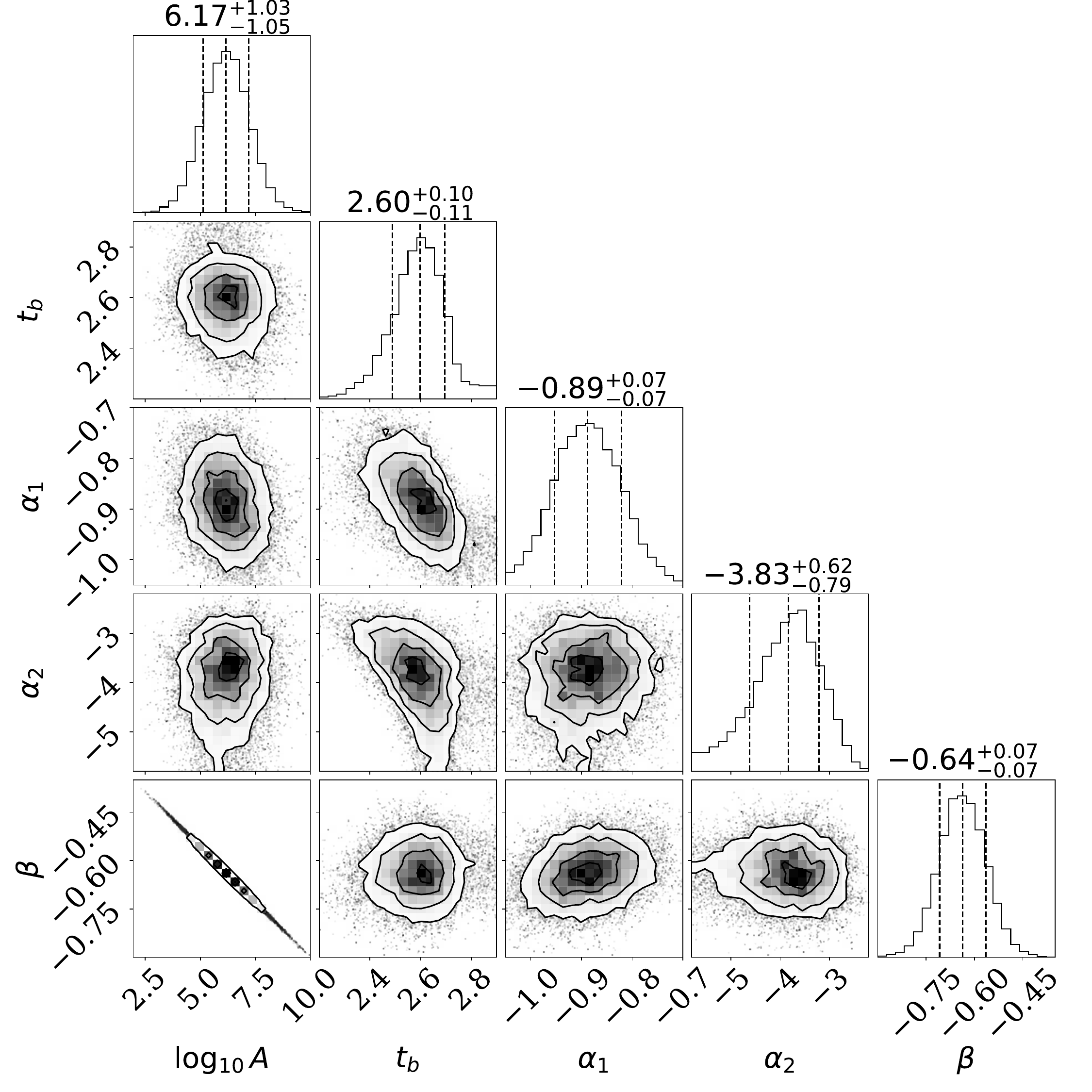}
 \caption{Corner plot of the phenomenological BPL fit to the afterglow of \target{}.  The $\alpha_2$ value constitutes a 3.6$\sigma$ tension with the post-break value predicted from closure relations, assuming the observed temporal break is the jet-break.}
 \label{fig:corner_bpl}
\end{figure}

\begin{figure}
 \includegraphics[width=\columnwidth]{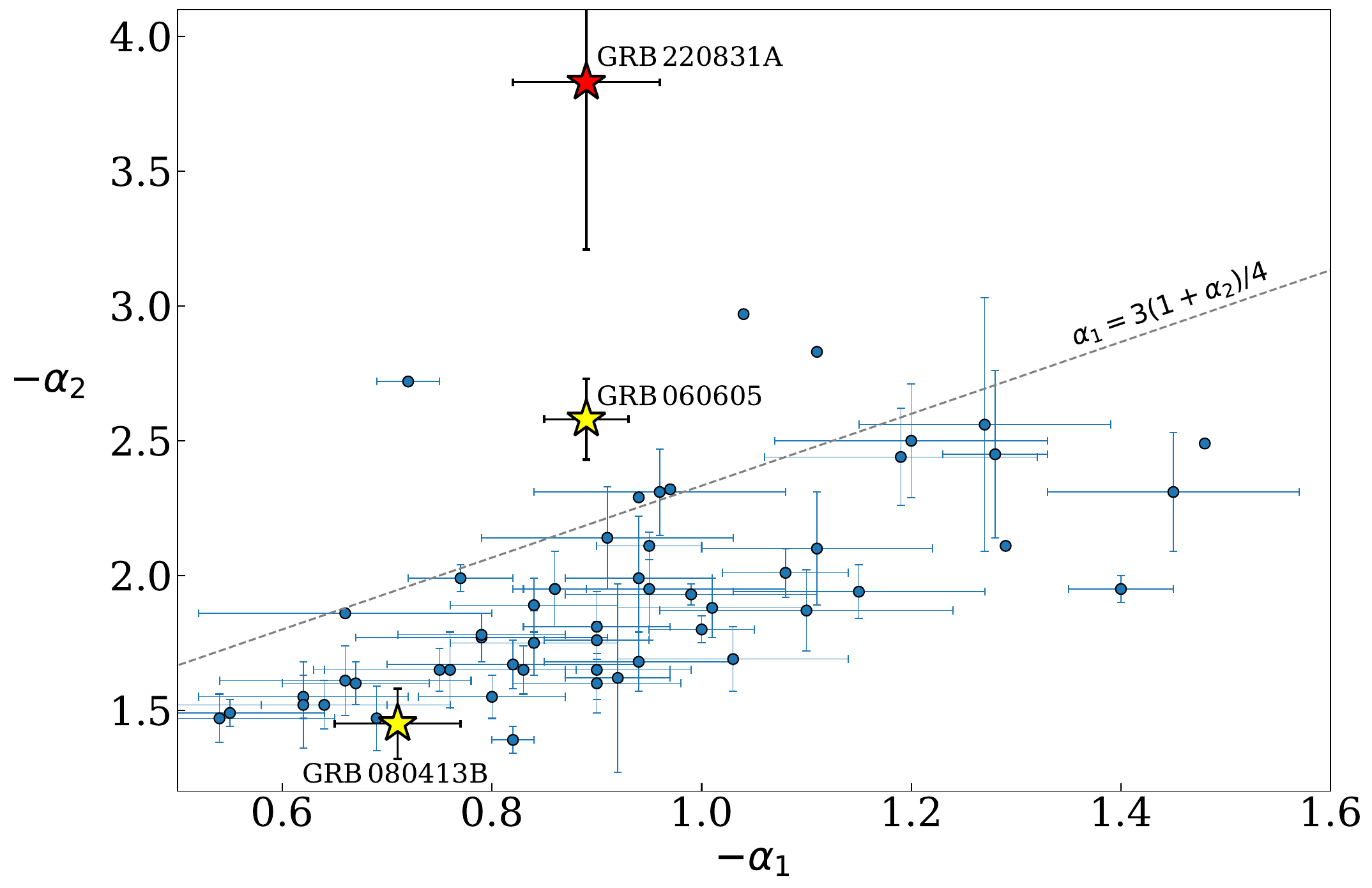}
 \caption{Temporal powerlaw indices for the sample of GRB afterglows analysed in \citet{GRBjetbreaks}.  We also plot \target{}'s location in parameter space and the predictions from forward shock closure relations, assuming a uniform circumburst density and $\nu_m < \nu < \nu_c$, denoted by the grey dotted line.  In the \citet{GRBjetbreaks} sample, the afterglows of GRBs 060605 and 080413B are plotted, both of which had a possible OIR internal plateau \citep{Li12}.  The values of $\alpha_1$ and $\alpha_2$ for GRB\,060605 are taken from \citet{GRB060605}.}
 \label{fig:breaks}
\end{figure}

From the pre-break values of $\beta$ and $\alpha_1$, we calculate $p=2.22 \pm 0.07$ \citep{afterglow_breaks}. This pre-break behavior, in terms of both temporal and spectral index, is consistent with the standard closure relations. However, if we interpret the sharp temporal break as a jet-break, the fitted value of $\alpha_2=-3.83^{+0.62}_{-0.79}$, is inconsistent with this prescription with a significance of 3.6$\sigma$, calculated from the posterior distribution shown in Figure \ref{fig:corner_bpl} \footnote{The posteriors of $\alpha_2$ are asymmetric, therefore, the errorbars in Figure \ref{fig:breaks} would not scale linearly with confidence level.}. Furthermore, it is not possible to generate such a decay with a forward shock.  The steepest decay possible would be limited by high latitude emission, resulting in $\alpha_2 = -2 + \beta$ \citep{KumarPanaitescu2000}.  It also constitutes an outlier in the distribution of observed temporal powerlaw indices, which is shown in Figure \ref{fig:breaks} \citep{GRBjetbreaks}.  Therefore, our inferred value of $\alpha_2$ is too steep to originate from a forward shock alone.  GRB\,060605 has a similarly steep post-break temporal powerlaw index \citep{GRB060605}, which can be seen in Figure \ref{fig:breaks}.  This will be discussed in Section \ref{sec:sec_emission} and Section \ref{sec:discussion}.

\subsubsection{Preliminary afterglow modeling}\label{sec:initial_fit}

In Figure \ref{fig:lightcurve}, we fit \target{}'s afterglow light curve using \textsc{afterglowpy}, a \textsc{Python} package for generating synthetic afterglow light curves from numerical forward shock models.  We assume a uniform density circumburst medium and a top-hat jet model with an on-axis viewing angle of 0$^{\circ}$.  We assume a top-hat as the presence of any angular structure would result in a shallower post-break light curve, leading to a larger discrepancy between forward shock models and our observations \citep[e.g.,][]{afterglowpy,BGG2020,BGG2022}.  We fit the data with the Bayesian inference library, \textsc{bilby} \citep{bilby}, with the nested sampler, \textsc{dynesty} \citep{dynesty}.  

Due to \target{}'s unconstrained progenitor and degeneracies in the redshift solution based on the Amati relation (see Figure \ref{fig:amati_relation}), we explore both a high ($z=2.4$) and low ($z=0.2$) redshift in our afterglow modelling.  Known degeneracies in $E_{K,\mathrm{iso}}$, $n_0$ and $z$ result in similar fits to the data.  As we favour a high-$z$, collapsar scenario, we assume $z=2.4$ for the remainder of this work, but present the result of the fits to the afterglow data, assuming $z=0.2$, in Appendix \ref{sec:z02_fits}.  

We fit all of the available data presented in Table \ref{tab:afterglowdata} for $E_{K,\mathrm{iso}}$, the jet-opening angle, $\theta_j$, $p$, the number density of the circumburst medium, $n_0$ and the forward shock's energy fraction stored in the electrons and the magnetic field, $\epsilon_e$ and $\epsilon_B$ respectively.  The resultant fit is shown in Figure \ref{fig:lightcurve} and the median of the posterior distribution has $\chi^2_{\mathrm{red}}=33.47$.  Given that our BPL fit to the OIR data yielded a value of $\alpha_2$ that was unusually steep it is unsurprising that the \textsc{afterglowpy} fails to simultaneously predict the post-break \textit{Gemini}/GMOS $i$-band detection and $r$-band upper limit at 4.6\,d post-burst and the pre-break OIR detections, particularly the VLT observations at $\sim$2.7\,d post-burst.  The best-fit model also underpredicts the observed flux density at 9\,GHz at approximately 11 days post-burst. At this observing frequency, scintillation-induced extrinsic variability may be as high as 100\% \citep[see][and references therein]{2020MNRAS.494.2449D}. With only one detection we cannot determine whether the observed flux density is representative of the intrinsic luminosity of the source, or whether it appears substantially brighter by chance. Alternatively it also could be due to the shortcomings of current forward shock models in simultaneously modeling X-ray, OIR and radio afterglows \citep[e.g.,][]{radio_afterglow_challenges}.  The parameter estimations from the resultant fit is shown in Table \ref{tab:forwardshock_params}.

\begin{table*}
    \caption{Forward shock parameter estimations from \textsc{afterglowpy} fits of \target{}'s afterglow assuming $z=2.4$. 
 $\epsilon_{\gamma}$ denotes the $\gamma$-ray efficiency of a given model and $\chi^2_{\mathrm{red}}$ is the reduced $\chi^2$ statistic for each model's fit to the data.  The corner plots associated with these fits are shown in Appendix \ref{sec:corner_plots}.}
    \label{tab:forwardshock_params}
    \centering
    \begin{tabular}{cccccccccc}
    \hline\hline
    Data & Fit &  $E_{K,\mathrm{iso}}$ & $n_0$ & $\theta_j$ & $p$ & $\log_{10} \epsilon_e$ & $\log_{10} \epsilon_B$ & $\epsilon_{\gamma}$ & $\chi^2_{\mathrm{red}}$ \\ 
         &     &  $\log_{10}$\,erg & $\log_{10}$\,cm$^{-3}$ & rad &  &  & &  & \\ \hline
    All & Forward shock & $52.51^{+0.76}_{-0.79}$ & $-2.43^{+1.13}_{-0.98}$ & $0.06^{+0.02}_{-0.02}$ & $2.27^{+0.06}_{-0.06}$ & $-0.66^{+0.51}_{-0.47}$ & $-1.85^{+1.17}_{-1.61}$ & $0.10^{+0.36}_{-0.09}$ & 33.47 \\
     & & & & & & & & \\ 
    OIR & Forward shock & $52.07^{+0.39}_{-0.12}$ & $-1.97^{+0.24}_{-1.16}$ & $0.07^{+0.01}_{-0.03}$ & $2.25^{+0.06}_{-0.06}$ & $-1$ & $-1$ & $0.26^{+0.10}_{-0.16}$ & 3.24 \\
     & & & & & & & & & \\
    OIR & Forward shock + & $52.18^{+0.18}_{-0.16}$ & $-2.41^{+0.52}_{-0.49}$ & $0.06^{+0.01}_{-0.01}$ & $2.27^{+0.07}_{-0.07}$ & $-1$ & $-1$ & $0.21^{+0.11}_{-0.10}$ & 2.00\\
    & flare & & & & & & &  \\
    OIR & Forward shock + & $52.05^{+0.19}_{-0.21}$ & $-2.75^{+0.56}_{-0.60}$ & $0.05^{+0.01}_{-0.01}$ & $2.27^{+0.07}_{-0.07}$ & $-1$ & $-1$ & $0.27^{+0.15}_{-0.12}$ & 1.11\\ 
    & internal plateau & & & & & & & & \\ \hline\hline
    \end{tabular}
\end{table*}

\subsection{Possible secondary emission components}\label{sec:sec_emission}

The deviations from the standard closure relations (see Section \ref{sec:multi_afterglow}) could be explained by invoking a separate emission component, dominating between approximately one and three days post-burst. Here we consider a few of the possible interpretations. As we cannot directly associate \target{} to a progenitor class (collapsar versus merger; see Section \ref{sec:prompt_analysis}) we consider both \textit{i}) a SN excess in Section \ref{sec:SN}, and \textit{ii}) a KN excess in Section \ref{sec:KN}. We further consider two progenitor independent mechanisms: \textit{iii}) a delayed flare in Section \ref{sec:flare} and \textit{iv}) an internal plateau in Section \ref{sec:internal_plateau}. 

\subsubsection{Supernova}\label{sec:SN}

LGRBs at low redshifts are typically found to have produced observable SN \citep{Hjorth2013,Cano2017}. The GRB-SN light curves peak at >10\,d post-burst in the rest-frame \citep{Hjorth2013,Cano2017}, which would result in a shallower observed value of $\alpha_2$ compared to a forward shock, which is the opposite of what we observe.

\begin{figure}
    \includegraphics[width=\columnwidth]{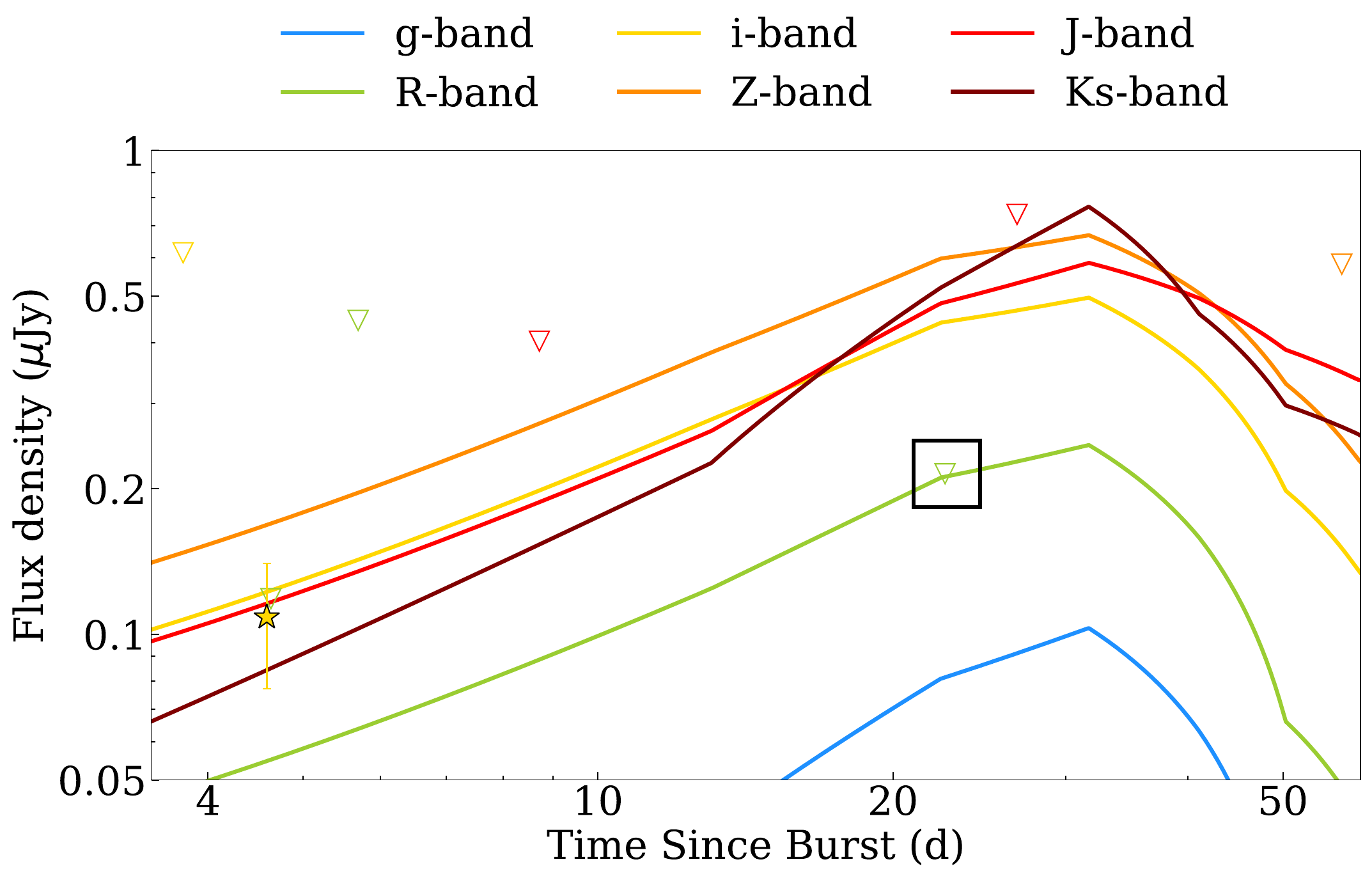}
    \caption{Model SN light curve based on observations of SN\,1998bw \citep{nugent_model,SNCosmo}, shifted to $z=0.86$.  We also show observations of \target{}'s OIR counterpart out to 57\,d post-burst which are listed in Table \ref{tab:afterglowdata}. In this scenario, the SN emission would dominate the \textit{Gemini}/GMOS $i$-band detection of the OIR counterpart at 4.6\,d post-burst and would be just below the VLT/FORS2 upper limit of $m_R>25.6$\,AB mag at 23\.d post-burst, which is highlighted in the plot with a black square.}
    \label{fig:SN_lightcurve}
\end{figure}

However, we can still use our sensitive OIR observations occurring out to 57 days post-burst, shown in Table \ref{tab:afterglowdata}, to constrain the presence of a SN accompanying \target{}.  We use a synthetic grid of light curves based on SN\,1998bw, the SN accompanying GRB\,980425 \citep{SN1998bw}, using the model described in \cite{nugent_model} with the \textsc{SNCosmo} library \citep{SNCosmo}. The $R$-band 5$\sigma$ upper limit of $m_R>25.6$\,AB mag at 23 days post-burst is most constraining for the existence of an SN and rules out synthetic light curves with $z\lesssim0.9$ (see Figure \ref{fig:SN_lightcurve}). This is consistent with the non-detection of an obvious host galaxy, which also implies a higher redshift (see Section \ref{sec:hostgalaxy}).

\subsubsection{Kilonova}\label{sec:KN}

If there was detectable KN emission associated with \target{}, we would expect reddening similar to that observed in other KNe.  However, the observed data prior to the temporal break is consistent with a constant spectral index, $\beta$.  Moreover, the steep decay of the $i$-band light curve would suggest the potential for a separate emission component that decays faster than the afterglow.  This evolution is not expected for KNe and was not observed in the KN accompanying GW170817, AT2017gfo \citep{at2017gfo1,at2017gfo2,at2017gfo3,at2017gfo4,at2017gfo5,at2017gfo6,at2017gfo7}, which decayed slower in $i$-band than typical, on-axis afterglow emission. Moreover, the lack of an obvious low-$z$ host disfavors a distance from which a KN would be detectable. We therefore conclude that the presence of a KN in our observations is unlikely.

\subsubsection{Flare}\label{sec:flare}

Most X-ray and optical afterglows, exhibit some flaring activity \citep{Li12,opticalflares,xrayflares}.  These flares are morphologically varied and have a range of different explanations.  Some can be explained by energy injection into the forward shock \citep{Burrows05,Romano06,Falcone06}, although many other explanations exist \citep[e.g.,][]{Dai06,Perna06,Duque22}. In the case of \target{}, the steep, post-break, powerlaw index would suggest an internal process over energy injection into the forward shock.  Optical flares like this have been observed before such as the optical afterglows to GRB\,070311 and GRB\,071010A \citep{Li12}.  Explanations for these flares include time-varying microphysics induced by a wind-bubble envirnoment \citep{Kong10} and the collision of internal shocks \citep{Guidorzi07}.

To test the existence of a flare in \target{}'s OIR counterpart, we describe the flare with a phenomenological smoothly broken powerlaw (SBPL) of the form in Equation \ref{eq:sbpl} for which we assume $\Delta=0.1$.  We simultaneously fit the SBPL and a forward shock model with \textsc{afterglowpy}, forcing a positive value for $\alpha_1$.  Our priors for the forward shock are based off a fit to the OIR data without the observations between 1 and 3 days post-burst.  This is because we interpret the data before 1 day post-burst and after 3 days post-burst to be dominated by the forward shock and those between 1 and 3 days to be dominated by a separate emission mechanism based off the results of the empirical fit in Section \ref{sec:obs_prop} forward shock fit in Section \ref{sec:initial_fit}.  

The forward shock fit assumes $\epsilon_e=\epsilon_B=0.1$ as they are consistent with our fit to the multi-wavelength data in Table \ref{tab:forwardshock_params} and, without fitting multi-wavelength data, these parameters are unconstrained.  We also assume that the flare has a spectral index of $\beta=-0.69$, consistent with the BPL fit to the OIR afterglow.  The data available is insufficient to constrain a change in the spectral energy distribution.  We also cannot constrain the secondary emission component's effects on the X-ray and radio observations due to the lack of sampling.  In this case, we therefore only fit to the OIR data to avoid making assumptions about secondary emission component's behaviour in other wavelengths.

\begin{figure*}
    \includegraphics[width=\textwidth]{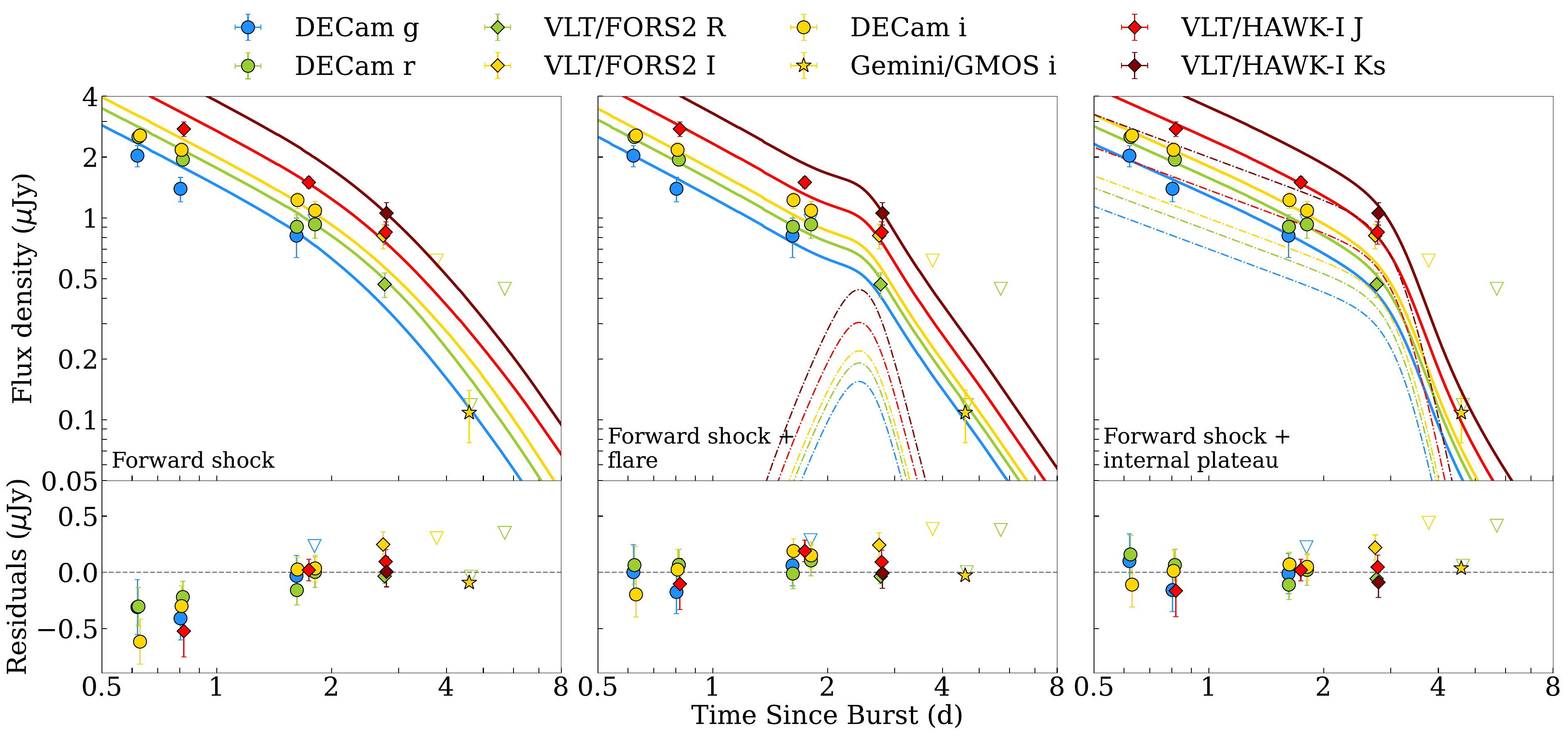}
    \caption{\textsc{afterglowpy} fits to \target{}'s OIR afterglow with additional emission components.  The solid lines denote the best-fit model, including both the forward shock and additional emission components. The dotted lines denote just the additional emission component.  The left-hand panel shows just forward shock emission, the central panel shows a flare, described by a SBPL starting at some time post-burst and the right-hand panel shows an internal plateau described by a BPL starting at the time of burst.  The best-fit parameters are listed in Tables \ref{tab:forwardshock_params} and \ref{tab:additional_params}.}
    \label{fig:lightcurve_models}
\end{figure*}

The parameters of the resultant forward shock and flare are presented in Table \ref{tab:forwardshock_params} and \ref{tab:additional_params} respectively and the light curve is shown in the central panel in Figure \ref{fig:lightcurve_models}.  Compared to a simple forward shock fit to the OIR data, we find moderate evidence for a flare with $\log_{10}\mathrm{BF}=0.96 \pm 0.07$, where BF is the Bayes factor.  We also note that, with the median of the posterior distribution, plotted in Figure \ref{fig:lightcurve_models}, there are still significant residuals, particularly with the $i$-band detections between 1 and 3\,d post-burst.
  
\begin{table*}
    \caption{Internal plateau and flare parameter estimations for fits to \target{}'s afterglow assuming $z=2.4$.}
    \label{tab:additional_params}
    \centering
    \begin{tabular}{cccccccc}
    \hline\hline
    Fit & $A$ & $t_0$ & $t_b$ & $\alpha_1$ & $\alpha_2$ & $\log_{10}\mathrm{BF}$ \\ 
         & $\log_{10}\mu$Jy & d & d &  &  & \\ \hline
    Forward shock + & $9.38^{+0.45}_{-2.59}$ & $0.68^{+0.63}_{-0.47}$ & $2.56^{+0.90}_{-0.42}$ & $2.78^{+4.43}_{-2.21}$ & $-6.16^{+2.88}_{-2.64}$ & $0.96 \pm 0.07$\\
    flare & & & & & \\
    Forward shock + & $9.44^{+0.22}_{-0.19}$ & $0$ & $3.21^{+0.51}_{-0.38}$ & $-0.70^{+0.39}_{-0.21}$ & $-9.36^{+3.96}_{-3.76}$ & $2.05 \pm 0.07$\\ 
    internal plateau & & & & & \\ \hline\hline
    \end{tabular}
\end{table*}

\subsubsection{Internal plateau}\label{sec:internal_plateau}

A significant fraction of both SGRB and LGRB X-ray afterglows have a plateau phase, where the decay of the light curve slows for a period of time before returning to a standard afterglow decay \citep{xrayplateaus}.  There are multiple interpretations for these external plateaus including energy injection into the forward shock from continued central engine activity \citep{Dalloso2011,xrayplateau_centralengine1,xrayplateau_centralengine2} and the result of angular structure in the jet \citep{EichlerGranot06,Oganesyan20,Beniamini2020}.

X-ray afterglows can also exhibit an `internal plateau' \citep{IP_Zhang2006,IP_Liang2006}, characterised by an extremely rapid decay in luminosity, such as observed, for example, from  GRB\,070110 \citep{GRB070110_discovery,IP_Beniamini2017}. Internal plateaus have also been proposed as explanation for the properties of the optical afterglows of GRBs 060605 and 080413B \citep{Li12}, though the evidence for a steep decay is not required by the data \citep[e.g.,][]{GRB060605}. 

The most popular of the proposed mechanisms to produce these internal plateaus is the presence of a long-lived millisecond magnetar central engine which spins down and eventually dissipates its energy by collapsing into a black hole \citep{IP_Lyons2010,IP_Rowlinson2010,IP_Zhang2014}.  This process would be unlikely to produce an OIR plateau. However, not all explanations require a long-lived central engine.  Another possible explanation proposed by \citet{IP_Beniamini2017} is photospheric emission arising from a moderately relativistic outflow (bulk Lorentz factor $\Gamma$\,$\lesssim$\,$20$) launched by the central engine (e.g., a black hole or magnetar) at a similar time as the highly relativistic material ($\Gamma$\,$\gtrsim$\,$100$) producing the forward shock. However, in this photospheric emission scenario it is still difficult to produce the OIR emission \citep[see][for details]{IP_Beniamini2017}.

Whilst the steep decay after the temporal break in \target{}'s OIR counterpart is similar to what is observed in internal plateaus.  To assess the possibility of an internal plateau in \target{}'s OIR light curve, we use a similar procedure as in Section \ref{sec:flare}.  We conduct a joint fit with a forward shock and a phenomenological SBPL, restricting $\alpha_1<0$ and $t_0=0$.  The resultant forward shock and SBPL parameters are shown in Tables \ref{tab:forwardshock_params} and \ref{tab:additional_params} respectively and the best-fit light curve is presented in the right panel of Figure \ref{fig:lightcurve_models}.  As expected, we obtain a comparatively steep post-break powerlaw index where $\alpha_1/\alpha_2=(7.5^{+9.2}_{-5.2})\times10^{-2}$.  This is typical of an internal plateau observed in X-rays \citep[e.g.,][]{IP_130831A,GRB070110_discovery,IP_Rowlinson2010} and is similar to the suspected OIR internal plateaus present in GRBs 060605 and 080413B \citep{Li12}.  With this fit, we find $\log_{10}\mathrm{BF}=2.05 \pm 0.07$, constituting strong evidence in comparison to a simple forward shock. We note that the lower number of free parameters and its contribution to the flux of the $t<1$\,d observations would naturally give it a larger Bayes factor than the flare explanation.  With the best-fit parameters presented in Table \ref{tab:additional_params}, we calculate a lower limit on the isotropic-equivalent energy release of the internal plateau of $>$10$^{50}$\,erg.

\section{Discussion}\label{sec:discussion}

\subsection{The origin of \target{}}

With the available data, we cannot assign a progenitor to \target{}.  The attempts to identify \target{}'s host galaxy and our study of its prompt emission have left the progenitor uncertain, with either a collapsar or a merger scenario open.  The narrow jet opening angle and low circumburst density we infer in Table \ref{tab:forwardshock_params} weakly favours a SGRB \citep{FongSGRBs,OConnor2020}.  However, there is a overlap in the circumburst densities \citep[e.g.,][]{Nysewander2009} and opening angles \citep[e.g.,][]{GRBjetbreaks,RoucoEscorial2022} among LGRBs and SGRBs.  We also acknowledge that the circumburst density measurements may be affected by the triple degeneracy in $E_{K,\mathrm{iso}}$, $n_0$ and $\epsilon_B$ that exists for bursts observed at $\nu_{\rm m}<\nu<\nu_{\rm c}$ \citep[e.g.,][]{degeneracy, degeneracy2}.  

Due to the lack of a significant association with any nearby galaxies (Figure \ref{fig:hostgal_field}), we favour a high-$z$ collapsar as the progenitor for \target{}.  This is also supported by the lack of a SN detection, which favours a distant origin at $z>0.86$.  Detecting a host galaxy for \target{} at $z=2.4$, would require a deep image with an upper-limit $m>27.2$\,AB mag which would correspond to $M>-18$ \citep{SchulzeGRBhosts}.  This is achievable with current 8-10\,m class ground-based and space-based facilities, which may be required to detect the host galaxy and determine its redshift.  

For future similar bursts, conducting rapid follow-up spectroscopy of their afterglow can yield a redshift measurement from absorption lines \citep[e.g.,][]{deUgartePostigo130603B,AguiFernandez160410A} and aid in the identification of a host galaxy and progenitor.  This is a significant challenge for SGRBs due to the comparative rareity of a bright OIR afterglow and their rapid fade rate, often fading below the spectroscopic detection threshold before a day post-burst.

For distinguishing merger-driven GRBs from collapsar-driven GRBs, the detection of SNe and KNe accompanying intermediate-class GRBs like \target{} can help understand the overlap between the populations of LGRBs and SGRBs.  Additionally, by conducting high-cadence searches for OIR afterglows \citep[e.g.,][]{ZTFdirtyfireballs,DWForphanafterglows} and the discovery of soft X-ray afterglows with facilities like \textit{Einstein Probe} \citep{EP,EP2022} can help understand the diversity, rate and angular structure of successful jets originating from BNS mergers and collapsars. These data would aid in tying emission properties to specific progenitors.

\subsection{\target{}'s unusual afterglow}

We observe departures from the standard forward shock closure relations in \target{}'s OIR afterglow at a few days after the burst (see Figure \ref{fig:lightcurve_models}).  We uncover strong evidence for an additional emission component which may be either a flare or internal plateau on top of the forward shock, based on the phenomenological SBPL fits we conducted.  However, the sampling of the light curve between 1 and 4 days post-burst is insufficient to discern the exact emission mechanism of the OIR excess.

As mentioned in Section \ref{sec:sec_emission}, there is evidence for internal plateau emission in the optical afterglows to GRBs 060605 and 080413B \citep{Li12}.  GRB\,060605's temporal powerlaw indices are shown in Figure \ref{fig:breaks} and exhibits a similar steepening in its evolution, albeit at earlier times, with a break $\sim$0.2\,d post-burst compared with \target{}'s $\sim$3\,d post-burst.  GRB\,060605's high redshift ($z=3.78$) means that this steepening is occurring very early in its rest frame evolution.  However, at $z=1.1$, GRB\,080413B has its temporal break at $\sim$1\,d post-burst (rest frame), a similar timescale to \target{}, if it is $z\gtrsim2$.  However, \citet{Filgas11} show that GRB\,080413B's optical afterglow is consistent with a two-component jet model, whereas the OIR counterpart to \target{}'s post-break powerlaw index is too steep for this prescription.

Furthermore, the short ($T_{90}<2$\,s) duration of \target{} presents a problem for the magnetar interpretation as a newly formed magnetar central engine would require $>10$\,s to impart an outflow with a large enough energy per baryon to produce a highly relativistic GRB \citep{Beniamini17}.  This problem can be mitigated, assuming a collapsar origin for \target{}, by requiring the time for the jet to break out of the stellar envelope to be $\sim$10\,s before quickly shutting off.  However, this scenario is statistically fine tuned and would require more evidence for us to favour it. Moreover, we note that it is not obvious from a theoretical perspective that an OIR internal plateau can be easily produced using the current theories. Through this lens, a delayed flare, originating from an internal process, is better explanation for our observations.

Consistent optical, hours-timescale cadence between 1 and 4 days post-burst would have better constrained the presence of a flare activity against an internal plateau by sampling its rise in brightness.  For bright OIR afterglows coordinated follow-up efforts by global telescope arrays such as the Gravitational Wave Optical Transient Observer \citep[GOTO;][]{GOTO}, the Las Cumbres Observatory \citep{LCO}, the Global Rapid Advanced Network Devoted to the Multi-messenger Addicts \citep[GRANDMA;][]{GRANDMA1,GRANDMA2} and Global Relay of Observatories Watching Transients Happen \citep[GROWTH;][]{GROWTH}, among others,  would aid in characterising brighter analogues of \target{}.

\section{Conclusions}\label{sec:conclusion}

The bimodal distribution of GRBs, comprising both short-hard and long-soft GRBs has typically been interpreted as being driven by binary neutron star mergers and collapsars respectively.  In recent years, observations of kilonovae associated with long-soft GRBs have shown the shortcomings of this assumption.  Exploring the parameter space between these two distributions may aid in constraining the fraction of collapsars in the short-hard regime and vice-versa.

In this work, we show that \target{} was an intermediate class GRB which shows a significant departure from a typical forward shock dominated afterglow.  We find that \target{} is observationally hostless which indicates that it is either a high redshift collapsar or a low redshift binary neutron star merger with a large angular offset from its host.  We favour a high-$z$ collapsar as the origin for \target{} due to \textit{i}) the prompt emission's comparative softness, \textit{ii}) its location on the Amati relation, \textit{iii}) the lack of significant association with nearby galaxies and \textit{iv}) the non-detection of an accompanying supernova or kilonova.

We fit two models to \target{}'s optical and near-infrared (OIR) counterpart, simulating a forward shock with both a delayed OIR flare and an OIR internal plateau, respectively.  We find that the addition of the phenomenological models allows us to fit the data better than a forward shock on its own.  However, the cadence of the OIR data is too slow to confirm the presence of an optical internal plateau over a flare or other additional emission mechanism.

For future GRB observations, we highlight the benefit of conducting rapid spectroscopic follow-up to obtain a redshift as it will allow for an easier determination of the event's progenitor.  We also identify the opportunity, by conducting high-cadence observations (e.g., hour cadence) of GRB OIR afterglows out to a few days post-burst, to constrain phenomena like central engine activity and long-lived magnetars.

\section*{Acknowledgements}

We thank Kathleen Labrie and the \textit{Gemini} External Helpdesk for helpful discussions regarding \textit{Gemini}/FLAMINGOS-2 data.

B.O. is supported by the McWilliams Postdoctoral Fellowship at Carnegie Mellon University. 

JC acknowledges funding by the Australian Research Council Discovery Project, DP200102102.

A.M. is supported by the Australian Research Council DE230100055.

A.C.G. and the Fong Group at Northwestern acknowledges support by the National Science Foundation under grant Nos. AST-1909358, AST-2308182 and CAREER grant No. AST-2047919. A.C.G. acknowledges support from NSF grants AST-1911140, AST-1910471 and AST-2206490 as a member of the Fast and Fortunate for FRB Follow-up team.

Parts of this research were conducted by the Australian Research Council Centre of Excellence for Gravitational Wave Discovery (OzGrav), through project numbers CE170100004 and CE230100016.

Research at Perimeter Institute is supported in part by the Government of Canada through the Department of Innovation, Science and Economic Development and by the Province of Ontario through the Ministry of Colleges and Universities

This work was supported by the European Research Council through the Consolidator grant BHianca (grant agreement ID 101002761) and by the National Science Foundation (under award number 12850).

Parts of this work was performed on the OzSTAR national facility at Swinburne University of Technology. The OzSTAR program receives funding in part from the Astronomy National Collaborative Research Infrastructure Strategy (NCRIS) allocation provided by the Australian Government, and from the Victorian Higher Education State Investment Fund (VHESIF) provided by the Victorian Government.

This work made use of data supplied by the UK \textit{Swift} Science Data Centre at the University of Leicester. Based on observations obtained at the international Gemini Observatory, a program of NSF's OIR Lab, which is managed by the Association of Universities for Research in Astronomy (AURA) under a cooperative agreement with the National Science Foundation on behalf of the Gemini Observatory partnership: the National Science Foundation (United States), National Research Council (Canada), Agencia Nacional de Investigaci\'{o}n y Desarrollo (Chile), Ministerio de Ciencia, Tecnolog\'{i}a e Innovaci\'{o}n (Argentina), Minist\'{e}rio da Ci\^{e}ncia, Tecnologia, Inova\c{c}\~{o}es e Comunica\c{c}\~{o}es (Brazil), and Korea Astronomy and Space Science Institute (Republic of Korea). Additionally, this work is based on data obtained from the ESO Science Archive Facility.

This research made use of \textsc{matplotlib}, a Python library for publication quality graphics \citep{matplotlib}, \textsc{SciPy} \citep{scipy}, \textsc{Astropy}, a community-developed core Python package for Astronomy \citep{astropy1,astropy2} and \textsc{scikit-learn}   \citep{sklearn}.

\section*{Data Availability}

The majority of the data used for this work is publicly available on data archives.  The data still in its proprietary period will be supplied upon reasonable request to the authors.



\bibliographystyle{mnras}
\bibliography{ref} 




\appendix

\section{A persistent artefact in FLAMINGOS-2 imaging}\label{sec:artefact}

In Section \ref{sec:data}, we report that some of the data collected with \textit{Gemini}/FLAMINGOS-2 at 9, 10 and 52\,d post-burst \citep[PI: O'Connor,][]{OconnorGCN} with the $J$ and $K_s$ filters suffers from an unknown image artefact likely introduced in the data reduction process (Kathleen Labrie, private communication), though we did not identify its exact cause in the data. Here, we assess that the detections in these filters are likely an image artefact, or, if it is astrophysical, that it is not associated with \target{}.  

There are three reasons for this. Firstly, additional late-time \textit{Gemini} observations in $J$-band, stacked between July 2023 and March 2024 also excluded this earlier detection as a host galaxy as they derive deeper, more sensitive limits, shown in Table \ref{tab:hostgaluplims}.

\begin{figure*}
    \includegraphics[width=\textwidth]{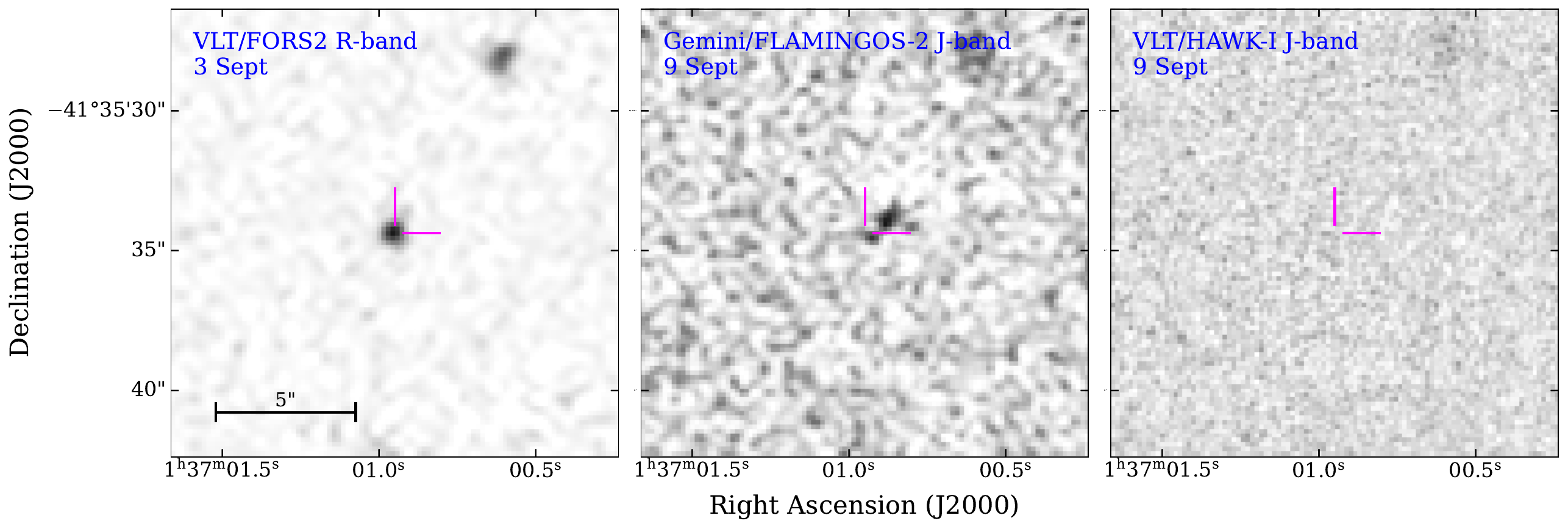}
    \caption{Images depicting the suspected image artefact in the \textit{Gemini}/FLAMINGOS-2 imaging. The left-hand panel shows the VLT/FORS2 $R$-band detection 3 September 2022, 2.7\,d post-burst, the central panel shows the \textit{Gemini}/FLAMINGOS-2 $J$-band $\sim24.5$\,AB mag detection of the suspected artefact, 9\,d post-burst and the right-hand panel shows the near contemporaneous VLT/HAWK-I $J$-band non-detection which had a 5$\sigma$ depth of $>24.9$\,AB mag. The images are not smoothed, but have been resampled with \textsc{SWarp} \citep{swarp}.}
    \label{fig:artefact}
\end{figure*}

Secondly, the \textit{Gemini} $J$ and $K_s$-band detections in question are spatially coincident with each other but offset from the rest of the OIR detections by $\sim$0.6\arcsec\, (Figure \ref{fig:artefact}).  Whilst this is a small offset if they  were due to a host galaxy, compared to the VLT afterglow detections with the same filters, we calculated a high significance of $\sim$14$\sigma$ for the deviation of the Gemini artefacts from \target{}'s localization after correcting for the astrometric tie uncertainty between the images. Given that they are not due to a host galaxy, this deviation rules out the association of them to \target{} as either an afterglow, kilonova, or supernova. Dust echoes, originating from a nearby GRB, could produce such a signature due to superluminal motion \citep{dust_echoes}.  However, this would occur hundreds of days post-burst, which is a much longer timescale than our observations at 9, 10 and 52\,d post-burst.

Lastly, near-simultaneous observations with VLT/HAWK-I of \target{} occurred within one hour after the initial \textit{Gemini}/FLAMINGOS-2 detection at $\sim$9\,d post-burst (Figure \ref{fig:artefact}).  To account for this non-detection, the source would likely have to fade by $>$0.4 AB mag in $\sim$1\,hr and subsequently rebrighten to account for the \textit{Gemini} $J$-band detection the next night at 10\,d post-burst.  This would be extremely peculiar behaviour for a GRB afterglow, SN, KN or a dust echo.

\section{Afterglow modeling assuming a low redshift}\label{sec:z02_fits}

In the main body of this work, we determined that \target{} is likely a $z>2$ collapsar based on the non-detection of a high significance host galaxy association, its location to the Amati relation, and upper-limits on SN emission.  However, we cannot strictly rule out a $z<0.4$ compact binary merger origin for \target{}.  In Sections \ref{sec:initial_fit} and \ref{sec:sec_emission}, we fit the data with various models, assuming $z=2.4$. Here, we provide fits, assuming the low-$z$ merger scenario with $z=0.2$. We note that while there is no obvious host galaxy associated to \target{}, a candidate host G0 does have a photometric redshift consistent with this redshift range but the probability of chance coincidence does not imply a robust association. 

\begin{figure*}
    \includegraphics[width=0.766\textwidth]{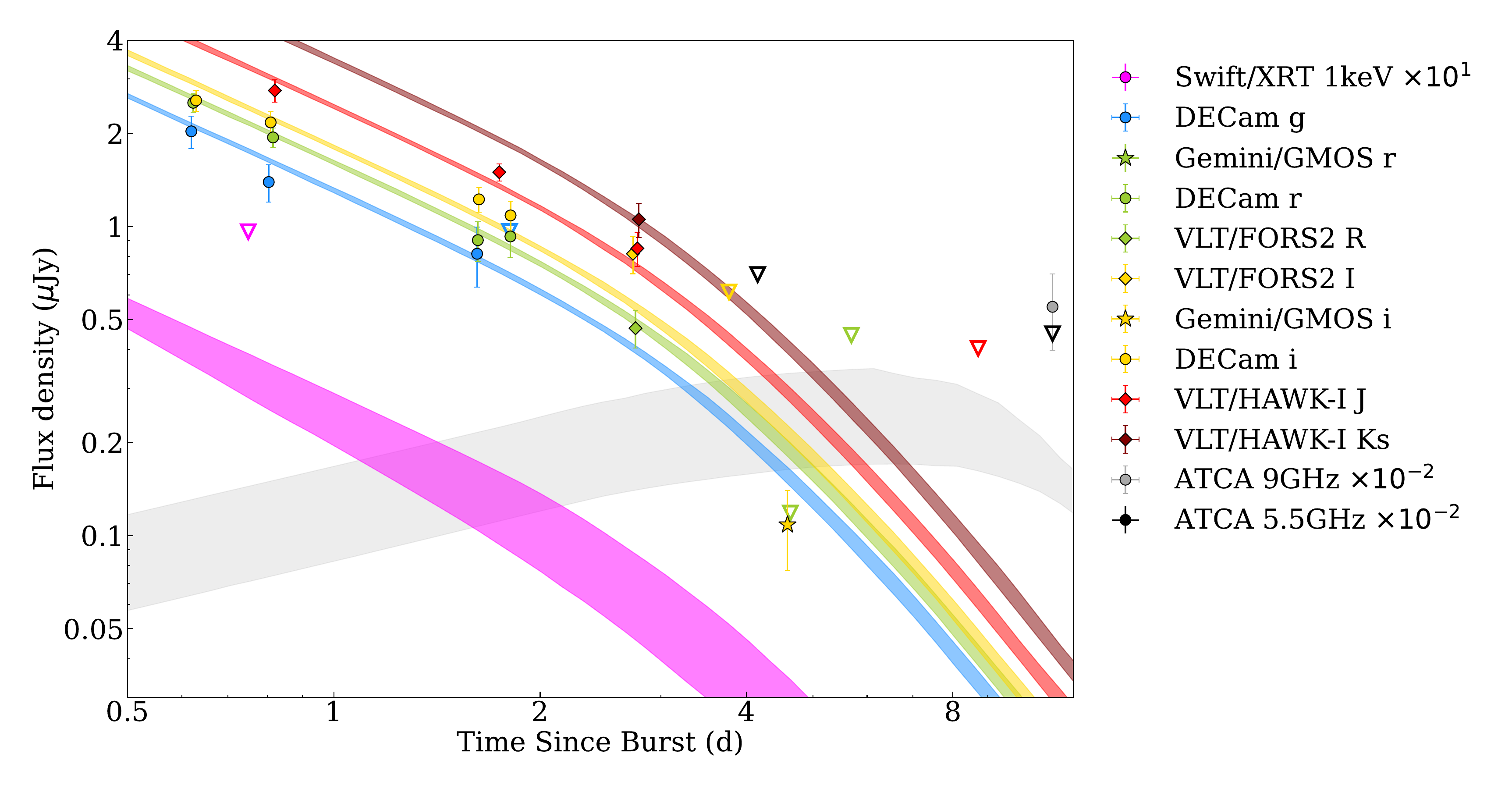}
    \caption{The light curve of \target{}'s afterglow.  The solid lines denote the best-fit \textsc{afterglowpy} \citep{afterglowpy} model, assuming $z=0.2$. The shaded regions show the 1$\sigma$ intervals for the fit.  The X-ray and radio light curves have been rescaled for better readability.}
    \label{fig:lightcurve_z02}
\end{figure*}

Figure \ref{fig:lightcurve_z02} shows the multi-wavelength light curve and \textsc{afterglowpy} fit to the multi-wavelength data.  The issues in fitting the steep temporal break at about 3 days post-burst, discussed in Section \ref{sec:initial_fit}, are still present regardless of whether \target{} is at $z=0.2$ or $z=2.4$.

In Tables \ref{tab:forwardshock_params_z02} and \ref{tab:additional_params_z02}, we find lower values of $E_{K,\mathrm{iso}}$ and lower values of $\chi_{\mathrm{red}}^2$ in for the OIR forward shock model and the forward shock and flare model.  Beyond a change to the energetics of the secondary emission component, it does not change our interpretation of the departures from closure relations discussed in Section \ref{sec:discussion}.

\begin{table*}
    \caption{Forward shock parameter estimations from \textsc{afterglowpy} fits of \target{}'s afterglow assuming $z=0.2$.}
    \label{tab:forwardshock_params_z02}
    \centering
    \begin{tabular}{cccccccccc}
    \hline\hline
    Data & Fit &  $E_{K,\mathrm{iso}}$ & $n_0$ & $\theta_j$ & $p$ & $\log_{10} \epsilon_e$ & $\log_{10} \epsilon_B$ & $\epsilon_{\gamma}$ & $\chi^2_{\mathrm{red}}$ \\ 
         &     &  $\log_{10}$\,erg & $\log_{10}$\,cm$^{-3}$ & rad &  &  & &  & \\ \hline
    All & Forward shock & $50.79^{+0.79}_{-0.76}$ & $-3.06^{+1.22}_{-1.50}$ & $0.12^{+0.09}_{-0.05}$ & $2.27^{+0.06}_{-0.06}$ & $-0.40^{+0.27}_{-0.36}$ & $-1.29^{+0.93}_{-1.50}$ & $0.08^{+0.32}_{-0.07}$ & 149.86
     \\
     & & & & & & & & \\ 
    OIR & Forward shock & $50.86^{+0.64}_{-0.52}$ & $-2.78^{+1.47}_{-1.78}$ & $0.12^{+0.10}_{-0.06}$ & $2.26^{+0.06}_{-0.06}$ & $-1$ & $-1$ & $0.83^{+0.12}_{-0.38}$ & 2.08 \\
     & & & & & & & & & \\
    OIR & Forward shock + & $51.09^{+0.09}_{-0.09}$ & $-3.49^{+0.25}_{-0.24}$ & $0.08^{+0.01}_{-0.01}$ & $2.27^{+0.06}_{-0.05}$ & $-1$ & $-1$ & $0.74^{+0.07}_{-0.11}$ & 1.43\\
    & flare & & & & & & &  \\
    OIR & Forward shock + & $50.99^{+0.10}_{-0.11}$ & $-3.63^{+0.28}_{-0.30}$ & $0.08^{+0.01}_{-0.01}$ & $2.26^{+0.07}_{-0.07}$ & $-1$ & $-1$ & $0.78^{+0.07}_{-0.16}$ & 1.03\\ 
    & internal plateau & & & & & & & & \\ \hline\hline
    \end{tabular}
\end{table*}

\begin{table*}
    \caption{Internal plateau and flare parameter estimations for fits to \target{}'s afterglow assuming $z=0.2$.}
    \label{tab:additional_params_z02}
    \centering
    \begin{tabular}{cccccccc}
    \hline\hline
    Fit & $A$ & $t_0$ & $t_b$ & $\alpha_1$ & $\alpha_2$ & $\log_{10}\mathrm{BF}$ \\ 
         & $\log_{10}\mu$Jy & d & d &  &  & \\ \hline
    Forward shock + & $9.50^{+0.36}_{-1.72}$ & $0.63^{+0.61}_{-0.45}$ & $2.52^{+0.68}_{-0.37}$ & $2.42^{+4.42}_{-1.94}$ & $-6.28^{+2.81}_{-2.53}$ & $1.97 \pm 0.07$\\
    flare & & & & & \\
    Forward shock + & $9.37^{+0.19}_{-0.18}$ & $0$ & $3.23^{+0.49}_{-0.39}$ & $-0.49^{+0.32}_{-0.28}$ & $-9.73^{+3.89}_{-3.53}$ & $2.05 \pm 0.07$\\ 
    internal plateau & & & & & \\ \hline\hline
    \end{tabular}
\end{table*}

\section{Results of Afterglow Modeling}\label{sec:corner_plots}

\begin{figure*}
    \includegraphics[width=0.66\textwidth]{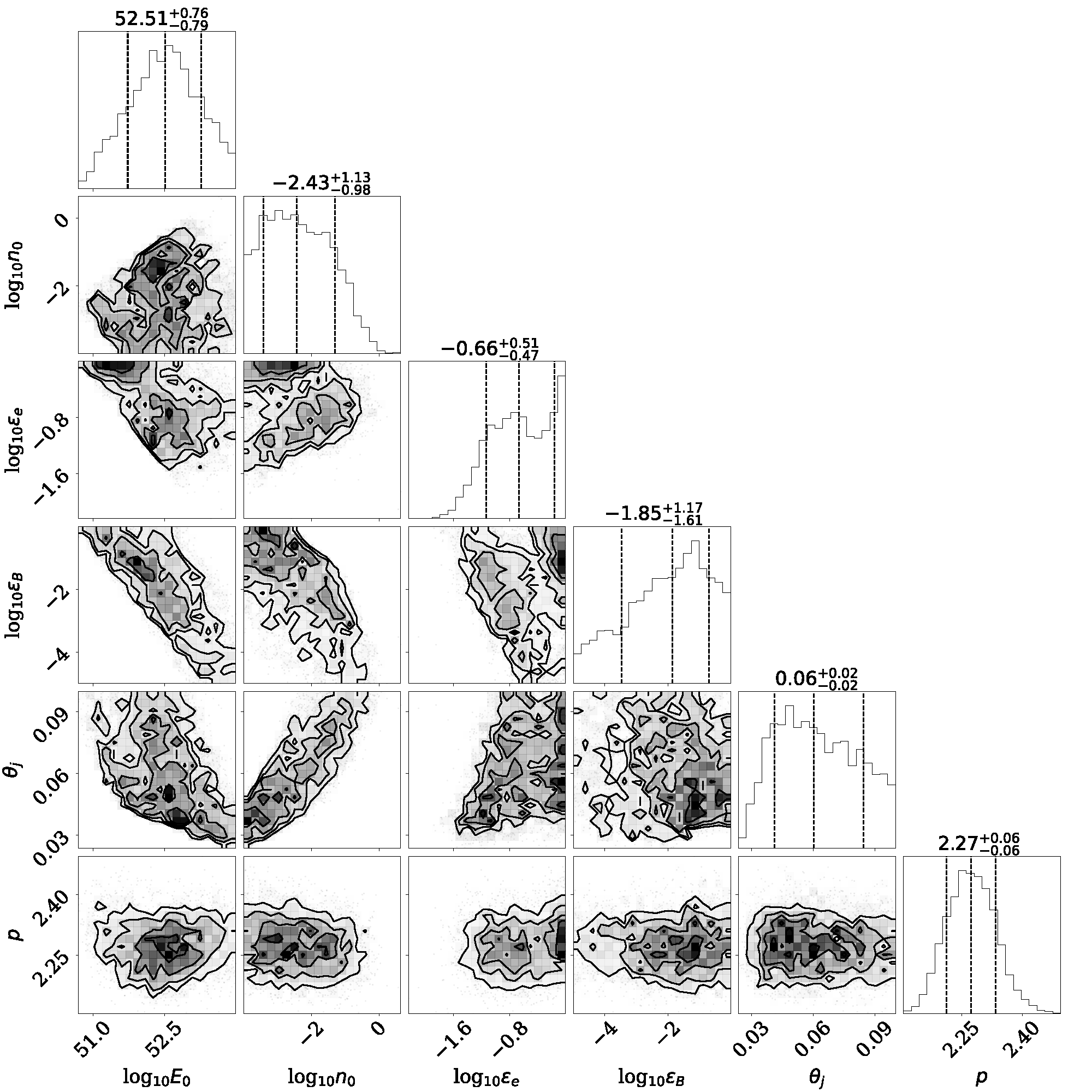}
    \caption{The corner plot corresponding to the multi-wavelength forward shock fit in Figure \ref{fig:lightcurve}.  We note the poor constraints on $\epsilon_e$ and $\epsilon_B$ which we attribute to the poor sampling of the X-ray and radio light curves.  $E_0$, $n_0$ and $\theta_j$ are in units of erg, cm$^{-3}$ and radians respectively.}
    \label{fig:cornerplot_multi}
\end{figure*}

\begin{figure*}
    \includegraphics[width=0.44\textwidth]{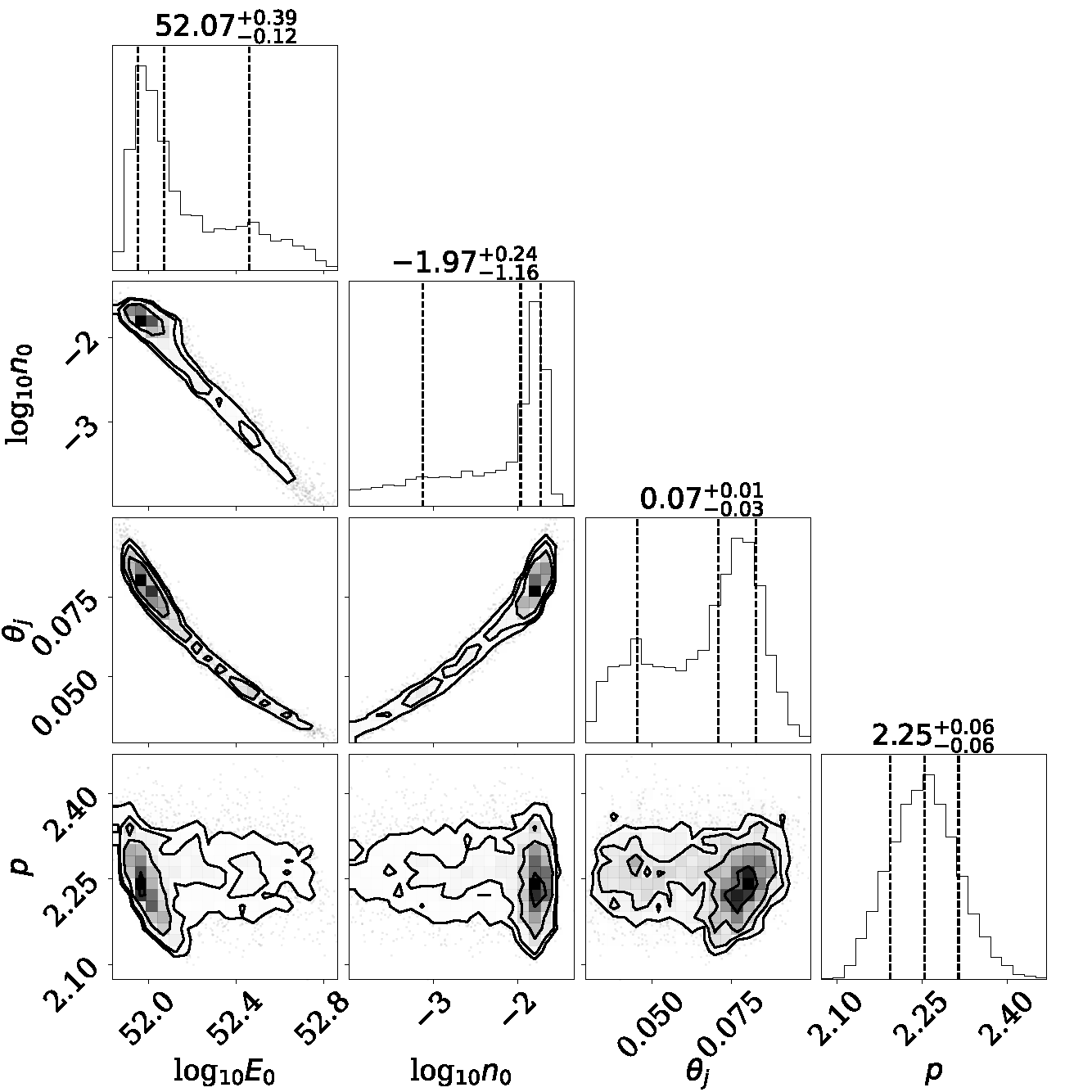}
    \caption{The corner plot corresponding to the forward shock fit in Figure \ref{fig:lightcurve_models}.}
    \label{fig:cornerplot_FS}
\end{figure*}

\begin{figure*}
    \includegraphics[width=\textwidth]{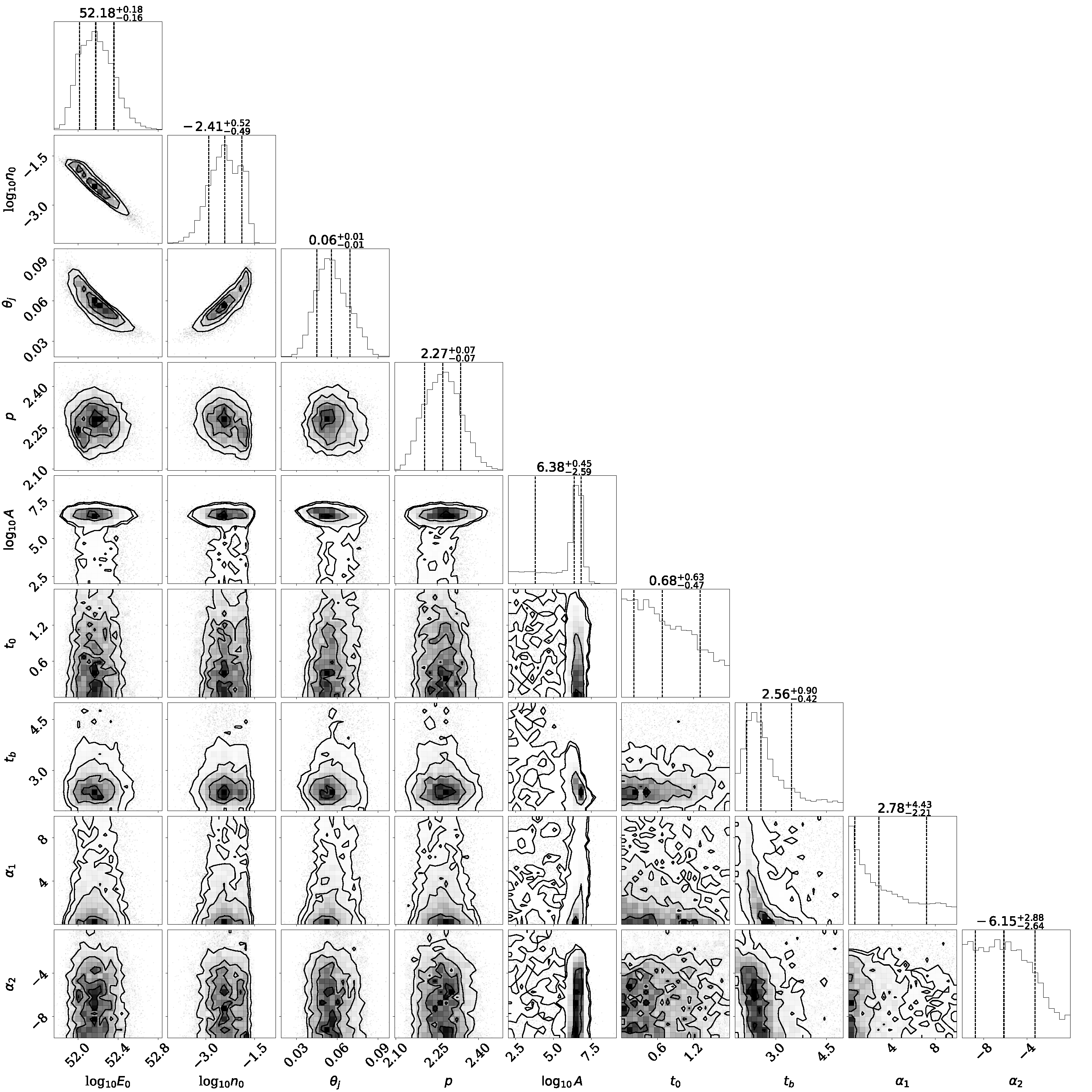}
    \caption{The corner plot corresponding to the forward shock and flare fit in Figure \ref{fig:lightcurve_models}.  $A$ is in units of mJy and $t_0$ and $t_b$ are in units of days.}
    \label{fig:cornerplot_flare}
\end{figure*}

\begin{figure*}
    \includegraphics[width=0.89\textwidth]{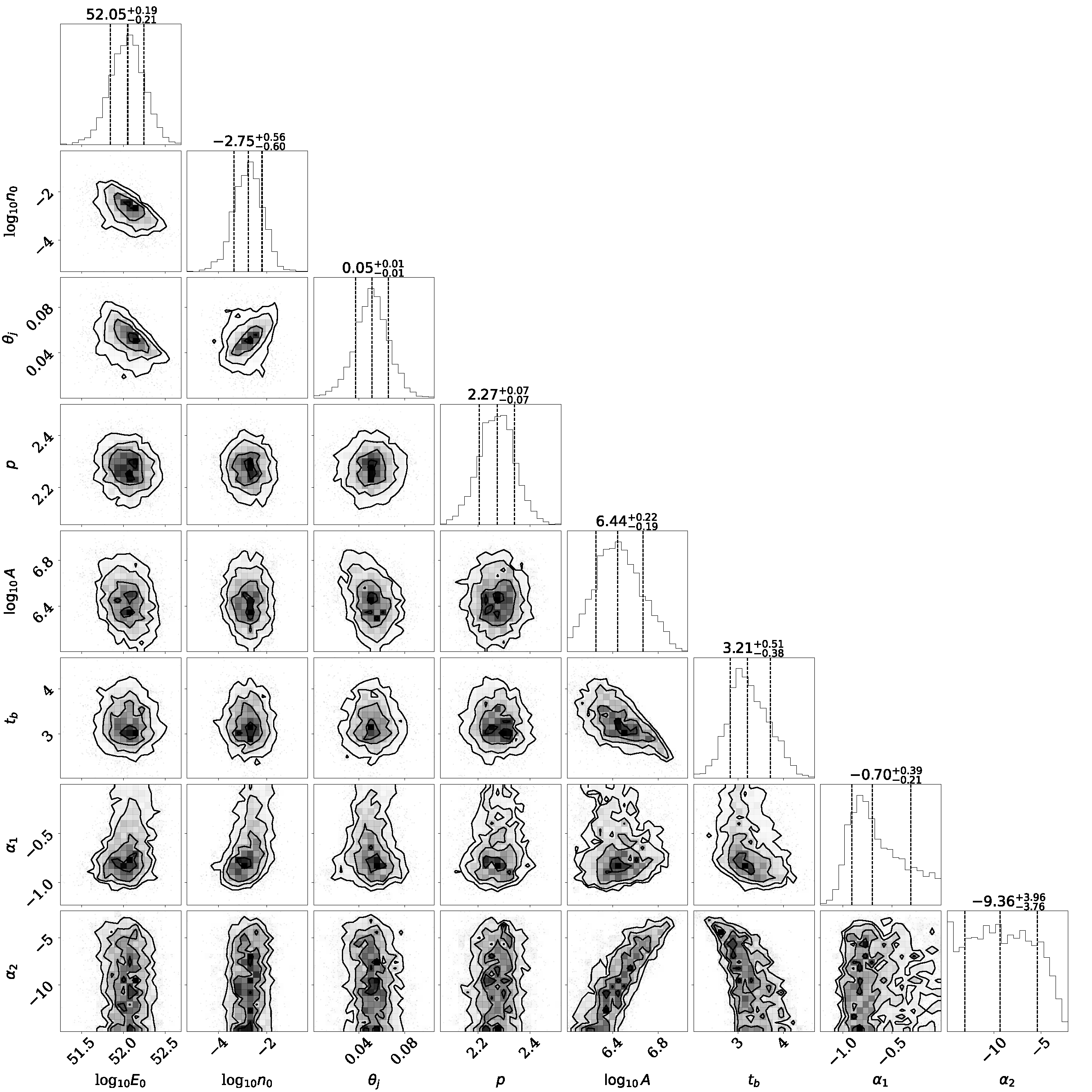}
    \caption{The corner plot corresponding to the forward shock and internal plateau fit in Figure \ref{fig:lightcurve_models}. }
    \label{fig:cornerplot_plateau}
\end{figure*}

In Section \ref{sec:multi_analysis}, we fit multiple models to \target{}'s multi-wavelength afterglow.  Here we present the corner plots from these fits.  This includes a forward shock fit to the entire multi-wavelength dataset in Figure \ref{fig:cornerplot_multi} and a forward shock fit to just the OIR data (Figure \ref{fig:cornerplot_FS}) and including a flare (Figure \ref{fig:cornerplot_flare}) and an internal plateau (Figure \ref{fig:cornerplot_plateau}).

\bsp	
\label{lastpage}
\end{document}